\documentclass[prb,aps,showpacs,groupedaddress,superscriptaddress,twocolumn,colorinlistoftodos]{revtex4-2}

\usepackage{amsmath,amssymb}
\usepackage[colorlinks=true,linkcolor=blue,urlcolor=blue,citecolor=blue]{hyperref}
\usepackage{nicefrac}
\usepackage{braket}
\usepackage{mathtools}
\usepackage{esvect}
\usepackage{amsthm}
\usepackage{fancyhdr}
\usepackage{amsfonts}
\usepackage{soul}
\usepackage{xcolor}
\DeclareMathOperator{\e}{e}

\newcommand{\plaqindex}{\ensuremath{j}}
\newcommand{\filling}{\ensuremath{\nu}}

\begin{document}
\title{Fate of chiral order and impurity self-pinning in flat bands with local symmetry}

\author{Maxime Burgher}
\affiliation{CENOLI,
Universit\'e Libre de Bruxelles, CP 231, Campus Plaine, B-1050 Brussels, Belgium}
\affiliation{International Solvay Institutes, Brussels, Belgium}

\author{Marco Di Liberto}
\affiliation{Dipartimento di Fisica e Astronomia ``G. Galilei'' \& Padua Quantum Technologies Research Center,
Universit\`a degli Studi di Padova, I-35131 Padova, Italy}
\affiliation{Istituto Nazionale di Fisica Nucleare (INFN), Sezione di Padova, I-35131 Padova, Italy}
\author{Nathan Goldman}
\affiliation{CENOLI,
Universit\'e Libre de Bruxelles, CP 231, Campus Plaine, B-1050 Brussels, Belgium}
\affiliation{International Solvay Institutes, Brussels, Belgium}
\affiliation{Laboratoire Kastler Brossel, Coll\`ege de France, CNRS, ENS-Universit\'e PSL, Sorbonne Universit\'e, 11 Place Marcelin Berthelot, 75005 Paris, France
}
\author{Ivan Amelio}
\affiliation{CENOLI,
Universit\'e Libre de Bruxelles, CP 231, Campus Plaine, B-1050 Brussels, Belgium}
\affiliation{International Solvay Institutes, Brussels, Belgium}

\date{\today}

\begin{abstract}
Interacting bosons on a single plaquette threaded by a $\pi$-flux can spontaneously break time-reversal symmetry, resulting in a chiral loop current.
Connecting such bosonic $\pi$-flux plaquettes in a dispersive configuration was recently shown to lead to long-range chiral  order.
Here, instead, we  design a chain of $\pi$-flux plaquettes that exhibits an all-flat-bands single-particle energy spectrum and an extensive set of local symmetries.
Using Elitzur's theorem, we show that these local symmetries prevent the emergence of long-range chiral order. 
Moreover, projecting the dynamics to a Creutz ladder model
with an effective intra-rung interaction
allows one to derive simple spin Hamiltonians capturing the ground state degeneracy and  the low-energy excitations, and to confirm the absence of chiral order.
Nevertheless, we show how to obtain gauge-invariant information from a mean-field approach, which explicitly breaks gauge-invariance.
Finally, we observe an ``impurity self-pinning'' phenomenon, when an extra boson is added on top of a ground state at integer filling, resulting in a non-dispersive density peak. Exact diagonalization benchmarks are also provided, and experimental perspectives are discussed.
\end{abstract}

\maketitle

\section{Introduction}
\label{sec:intro}

Spontaneously broken time-reversal symmetry has emerged in the last decades as one of  the most intriguing paradigms in condensed matter physics~\cite{simon2002,Neupert2021,Li2021,Cai2023,Park2023,Zeng2023,Liu2006, Wirth2011,DiLiberto2023}. 
For instance, the interplay of quantum geometry~\cite{Torma2022}, weak dispersion and interactions has been shown to stabilize the superconducting and Chern insulating phases observed in twisted bilayer graphene~\cite{cao2018super,cao2018correl},
and to give rise to the anomalous integer~\cite{Li2021} and fractional~\cite{Cai2023,Park2023,Zeng2023} Chern insulators  recently discovered in two dimensional materials.
Atoms loaded in $p$-bands were shown to spontaneously break time reversal symmetry and form chiral superfluids~\cite{Liu2006, Wirth2011}, a mechanism that can be reproduced in the lowest band of dimerized lattices with $\pi$-flux ~\cite{DiLiberto2023}. Aharonov-Bohm interference due to $\pi$-flux is known to lead to caging and localization \cite{Vidal2000} and has been demonstrated in photonic platforms, both with
 coupled waveguide arrays \cite{Mukherjee2018, Ozawa2019} and  superconducting circuits~\cite{martinez2023flatband}.

A conceptually appealing limit where to study the interplay of complex hopping phases,  quantum geometry and interactions is provided by models featuring perfectly flat bands. In this context, the emergence of superconducting transport~\cite{peotta2015,huhtinen2022revisiting} and  Bose condensation of pairs~\cite{Huber2010,Takayoshi2013,
Tovmasyan2013,Tovmasyan2018,
Julku2021_quantum_geometry,Julku2021} were demonstrated, in spite of the lack of single-particle dispersion.
In particular, it is important to recall that all-flat band systems come with an extensive set of local symmetries, related to the parity of local Wannier occupations. Since the condensation order parameter is odd under parity, and, as prescribed by Elitzur's theorem~\cite{Elitzur1975},   local symmetries cannot be spontaneously broken, single-particle condensation is forbidden. 
Hopping of pairs, instead, preserves parity locally, and pair condensation is allowed~\cite{Tovmasyan2018}.

In this work, we consider a different symmetry breaking scenario leading to chiral ground states, and we investigate the fate of such chiral order in an all-flat band array configuration.
As studied in Ref. \cite{DiLiberto2023}, for a single plaquette pierced by a $\pi$-flux and loaded with a large number of bosons, interactions can lead to  the spontaneous breaking of time-reversal symmetry. 
This manifests itself in chiral currents flowing clockwise or counterclockwise around the plaquette.
Furthermore, when several plaquettes are connected to form dispersive Bloch bands using a configuration analogous to the so-called BBH model~\citep{Benalcazar2017},  chiral long-range order of the Ising type emerges.

On the contrary, in our all-flat band configuration, we demonstrate that  Elitzur's theorem prevents any chiral correlation at finite distance.
This 
is also illustrated by resorting to projected effective models, including a Creutz ladder~\cite{Creutz1999} with unconventional interactions. 
We also discuss how to obtain gauge invariant information from mean-field theory, which explicitly breaks the local symmetries.
These results provide conceptual insights on the relation between band flatness and symmetry breaking.
Moreover, we observe an intriguing ``impurity self-pinning'' phenomenon:
when an extra boson is added with respect to an  integer filling configuration, a non-dispersive, exponentially localized density peak is supported in the ground state. While the created pair should in principle be dispersive, it is instead energetically bound to a
self-induced attractive potential in our case. 
We remark that this is
not the consequence of a many-body Aharonov-Bohm caging, a situation well studied in the literature~\cite{Tovmasyan2018,Danieli2020}.

The paper is structured as follows. In Sec.~\ref{sec:model} we introduce the model, perform the projection to a Creutz ladder with angular momentum dependent interactions and  establish the local symmetries.
In Sec.~\ref{sec:chiral_orbital_order} we review Elitzur's theorem and demonstrate the impossibility of spatial chiral order in our model; effective spin Hamiltonians are derived in specific regimes, confirming the above conclusion. The role of gauge invariance at the mean-field level is investigated.
The impurity self-pinning   is studied in Sec.~\ref{sec:pair_superfluidity},
using both exact diagonalization and the Wannierized Hamiltonian of the lowest single-particle band.
Finally, we draw our conclusions in Sec.~\ref{sec:conclusion}.

\section{Model, projection and symmetries}
\label{sec:model}

In this section, we introduce our  model, a specific array of $\pi$-flux plaquettes filled with interacting bosons, and discuss the  regime where a projection to a Creutz ladder is justified, allowing to halve the number of degrees of freedom. The local symmetries of the (full) plaquette and (projected) ladder models are explained. 
The higher-dimensional 
generalization of the model is presented in Appendix \ref{ap:2D}.
\subsection{Model}
\label{subsec:intro}

\begin{figure}[t]
    \centering
    \includegraphics[scale=0.4]{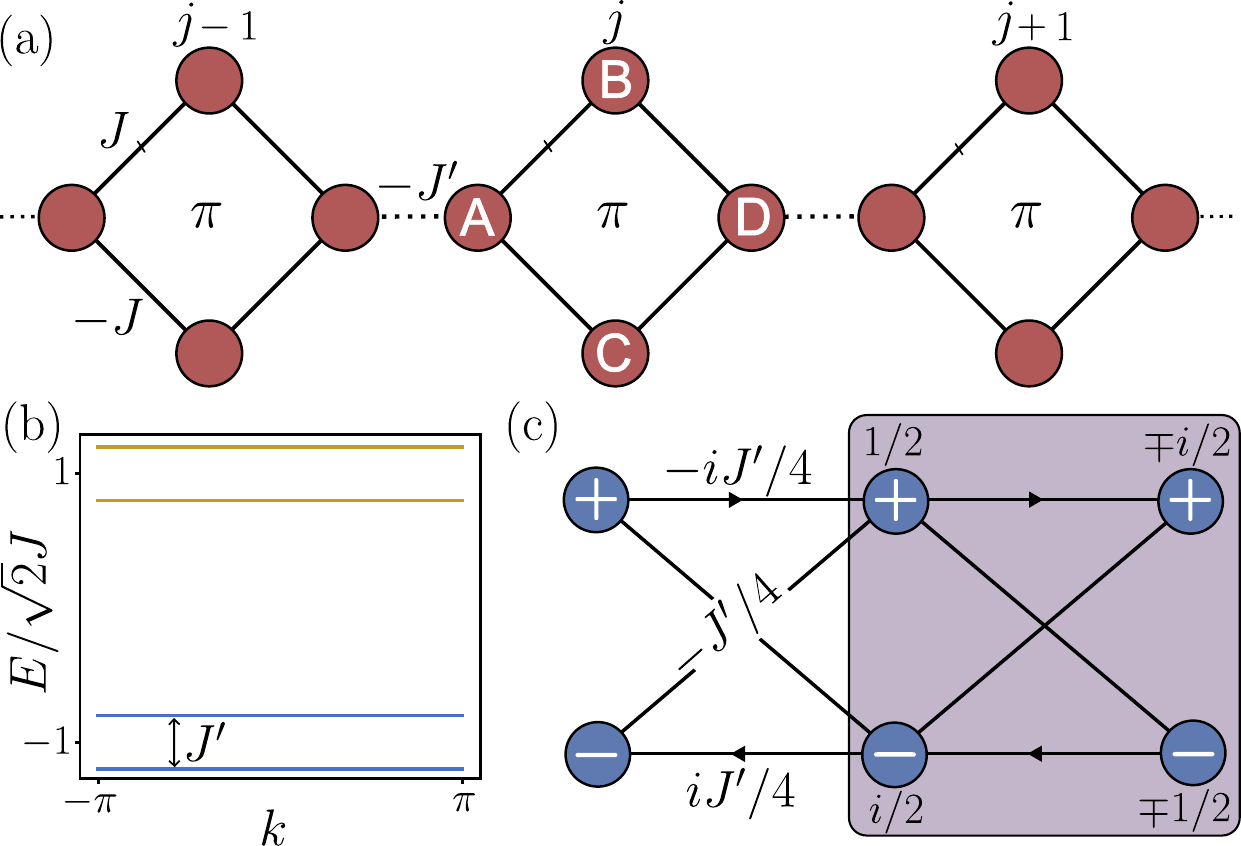}
    \caption{(a) Array of $\pi$-flux plaquettes, each consisting of four sites $A,B,C$ and $D$. In our gauge, the phase $\pi$ is concentrated on the link between $A$ and $B$ sites. (b) Single-particle spectrum displaying four completely flat bands. In the limit $J \gg U,J'$,  we perform a projection to the low-energy subspace consisting of the two blue bands. (c) 
    The kinetic part of the projected Hamiltonian in the angular-momentum basis, see Eq. (\ref{eq:H0_proj}),  can be represented as a Creutz ladder. The compact Wannier  orbitals  of both bands are illustrated in the purple box, with their respective  amplitudes from Eq. (\ref{eq:wannier_operator}).}
    \label{fig:model}
\end{figure}

In the following, we consider a chain of $L$ $\pi$-flux plaquettes connected in a non-dispersive way.
Each plaquette $\plaqindex$ consists of four inequivalent sites labeled by 
$\sigma=A,B,C,D$, as in Fig.~\ref{fig:model}(a). 
Bosons are created by the operator $\hat{\alpha}_{\sigma, \plaqindex}^\dagger$, or, within a parallel notation, by 
$\hat{a}_\plaqindex^\dagger = \hat{\alpha}_{A, \plaqindex}^\dagger , \hat{b}_\plaqindex^\dagger = \hat{\alpha}_{B, \plaqindex}^\dagger, \hat{c}_\plaqindex^\dagger = \hat{\alpha}_{C, \plaqindex}^\dagger$ and $\hat{d}_\plaqindex^\dagger = \hat{\alpha}_{D, \plaqindex}^\dagger$.
The Hamiltonian of  the $\plaqindex$-th plaquette is given by 
\begin{multline}
    \hat{H}^{(\plaqindex)} =  -J (\text{e}^{i\pi} \hat{a}_\plaqindex^\dagger \hat{b}_\plaqindex + \hat{a}_\plaqindex^\dagger \hat{c}_\plaqindex + \hat{b}_\plaqindex^\dagger \hat{d}_\plaqindex + \hat{c}_\plaqindex^\dagger \hat{d}_\plaqindex + H.c.)
    +
    \\
    + \frac{U}{2} \sum_\sigma \hat{n}_{\sigma, \plaqindex}
    (\hat{n}_{\sigma, \plaqindex}-1),
\end{multline}
where $J$ is the hopping constant and 
$U \geq 0$ the strength of the on-site repulsive interaction. The density operator is  $\hat{n}_{\sigma, \plaqindex} = \hat{\alpha}_{\sigma, \plaqindex}^\dagger \hat{\alpha}_{\sigma, \plaqindex}$.
Crucially, the hopping between sites $A$ and $B$ occurs with an extra minus sign, and is responsible of the Aharonov-Bohm destructive interference induced by the insertion of a $\pi$-flux.

 We now introduce extra  links with hopping constant $J'$ to connect the $D$ site of one plaquette with the $A$ site of the successive one, as sketched in Fig.~\ref{fig:model}(a) by the dotted lines. 
In summary, the overall Hamiltonian reads
\begin{equation}
     \hat{H} = \sum_{\plaqindex=1}^L 
     \left[
     \hat{H}^{(\plaqindex)}
     -J' ( \hat{d}_{\plaqindex}^\dagger \hat{a}_{\plaqindex+1}
     + \hat{a}_{\plaqindex+1}^\dagger \hat{d}_{\plaqindex})
     \right].
    \label{eq:H}
\end{equation}

In the non-interacting case, $U=0$, the Hamiltonian 
is diagonalized by means of Bloch theorem.
As illustrated in Fig.~\ref{fig:model}(b), this yields a single-particle spectrum consisting of four perfectly flat bands  of  energies 

\begin{equation}
    \begin{split}
        \varepsilon_{0} = - \varepsilon_{3} =
        & -\frac{1}{2}\Big(  \sqrt{8J^2 + J'^2} + J'\Big), 
        \\
        \varepsilon_{1} = - \varepsilon_{2} =
        & -\frac{1}{2}\Big( \sqrt{8J^2 + J'^2} - J'\Big).
    \end{split}
\end{equation} 

In the following we will be interested in the regime where $J \gg U, J'$.
In this case, the gap between the two lowest bands equals $\varepsilon_{1} -  \varepsilon_{0} = J'$ and the  gap  between these two and the other two bands is of the order of $\varepsilon_{2} -  \varepsilon_{1} \sim 2\sqrt{2}J$. We will then focus on the physics of the low-energy subspace obtained by projecting out the two high-energy bands.

\subsection{Projection in the angular-momentum basis}
\label{subsec:projection}

In order to study the low-energy physics in the regime $J \gg U, J'$, it is convenient to review the physics of a single plaquette.
Because of the Aharanov-Bohm phase and the resulting destructive interference, the single-particle spectrum of a plaquette features two degenerate doublets at energies $\pm 2\sqrt{2} J$. 
Focusing on the lowest energy single-particle doublet, it turns out that one can choose the eigenstates to be orthonormal and uniform in space, and we call the corresponding creation operators  $\beta_{+,\plaqindex}^\dagger$ and $\beta_{-,\plaqindex }^\dagger$.
The explicit expressions read
$\hat{\beta}_{\pm, \plaqindex}^\dagger
=
\sum_{\sigma} \phi_\pm(\sigma) \hat{\alpha}_{\sigma, \plaqindex}^\dagger
=
\frac{1}{2} \left(
\frac{1 \mp i}{\sqrt{2}} \hat{a}_\plaqindex^\dagger \pm i \hat{b}_\plaqindex^\dagger
+ \hat{c}_\plaqindex^\dagger
+ \frac{1 \pm i}{\sqrt{2}} \hat{d}_\plaqindex^\dagger
\right)
$.
The plus and minus subscripts are assigned noticing that these two states are eigenstates of 
the angular-momentum operator
\begin{equation}
    \hat{\mathcal{L}}_{z,\plaqindex} = \frac{-i}{\sqrt{2}}(\hat{a}^\dagger_\plaqindex \hat{c}_\plaqindex + \hat{c}^\dagger_\plaqindex \hat{d}_\plaqindex + \hat{d}^\dagger_\plaqindex \hat{b}_\plaqindex - \hat{b}^\dagger_\plaqindex \hat{a}_\plaqindex) + \text{H.c.},
    \label{eq:loop-current}
\end{equation}
with eigenvalue
$\pm 1$, respectively. Note that $\hat{\mathcal{L}}_{z,\plaqindex}$ is constructed as the loop-current operator around the $\plaqindex$-th
plaquette (divided by a $\sqrt{2}$ factor).
Physically, the plus (minus) sign of the eigenstates corresponds to a chiral current flowing clockwise (counterclockwise, respectively).

The spatial uniformity of the $\pm$ orbital densities  makes these states a convenient starting point to include interactions. 
In other words, the fact that $|\phi_\pm(\sigma)|^2 = \frac{1}{4} ~ \forall \sigma$
entails that, at the mean-field level, the $\pm$ orbitals optimize the repulsive interactions, in contrast to some arbitrary superposition (which at the non-interacting level would be degenerate).
In the weakly interacting regime $J \gg U$ and in the sector with $\filling$ bosons in the $\plaqindex$-th plaquette, approximate degenerate ground states of 
$\hat{H}^{(\plaqindex)}$
are then provided by $|+\rangle_\plaqindex =\frac{1}{\sqrt{\filling!}} (\beta_{+,\plaqindex}^\dagger)^\filling |0\rangle$ and $|-\rangle_\plaqindex = \frac{1}{\sqrt{\filling!}} (\beta_{-,\plaqindex}^\dagger)^\filling |0\rangle$, where $|0\rangle$ is the vacuum state. This physics was investigated in \cite{DiLiberto2023}, and the take-home message is that, at the many-body level, the interaction favors chiral correlations.
We mention that similar physics is obtained when bosons are loaded in $p$-bands \cite{Liu2006, Pinheiro2012, Collin2010, Li2012}.

When different plaquettes are connected, the  orbitals associated with $\hat{\beta}_{+,\plaqindex}^\dagger, \hat{\beta}_{-,\plaqindex}^\dagger$ provide a convenient basis to perform the projection onto the low-energy manifold in the regime $J \gg U, J'$.  
In this case, higher energy single-plaquette orbitals can be neglected, while the interaction and inter-plaquette hopping  mix the  $\hat{\beta}_{+,\plaqindex}^\dagger, \hat{\beta}_{-,\plaqindex}^\dagger$.
We note that the projection operator is strictly local, as can be proven based on general arguments~\cite{Sathe2021}.
Upon projection,  $J$ is only responsible for an energy shift of $\sqrt{2}J$ per particle, which is irrelevant when focusing on sectors with a given particle number. Leaving out this chemical potential, the Hamiltonian becomes

\begin{equation}
    \hat{H}^{\rm proj}
    =
    \hat{H}^{\rm proj}_{DA}
    +
    \hat{H}^{\rm proj}_{\text{int}},
    \label{eq:Hproj_total}
\end{equation}
where the interaction term reads
\begin{equation}
    \hat{H}^{\rm proj}_{\text{int}} = \sum_\plaqindex  \frac{3U}{16} \hat{n}_{\plaqindex}^2 - \frac{U}{16} \hat{L}_{z,\plaqindex}^2 -\frac{U}{8} \hat{n}_{\plaqindex},
    \label{eq:Hint_proj}
\end{equation}
with the plaquette density $\hat{n}_\plaqindex = \hat{\beta}^\dagger_{+,\plaqindex}\hat{\beta}_{+,\plaqindex} + \hat{\beta}^\dagger_{-,\plaqindex}\hat{\beta}_{-,\plaqindex}$ and the projected angular momentum operator $\hat{L}_{z,\plaqindex} = \hat{\beta}^\dagger_{+,\plaqindex}\hat{\beta}_{+,\plaqindex} - \hat{\beta}^\dagger_{-,\plaqindex}\hat{\beta}_{-,\plaqindex}$. The latter is related to the loop-current operator of Eq. \eqref{eq:loop-current} via $\hat{\mathcal{L}}^{\text{proj}}_{z,\plaqindex} = \hat{L}_{z,\plaqindex}$, making a direct link with the chirality of a plaquette.
Also, notice that these interactions are local in $\plaqindex$, but nonlocal within a rung (i.e. the unit cell containing the $\pm$ orbitals). Crucially, they favor a 
uniform distribution of particles between plaquettes, but an inhomogeneous concentration on the $+$ or $-$ orbitals within  individual plaquettes.
This unbalance reflects the maximization of orbital angular momentum and the onset of chiral order anticipated above. The effect of a similar angular-momentum interaction was discussed in Ref. ~\cite{Junemann2017} for a fermionic Creutz ladder.

The term $\hat{H}_{DA}$
of the Hamiltonian that describes inter-plaquette hopping is projected to
\begin{equation}
    \begin{split}
        \hat{H}_{DA}^{\text{proj}} =& \sum_\plaqindex - \frac{J'}{4}  \Big[i\hat{\beta}_{+,\plaqindex+1}^\dagger \hat{\beta}_{+,\plaqindex} + \hat{\beta}_{+,\plaqindex+1}^\dagger \hat{\beta}_{-,\plaqindex} \\
        & + \hat{\beta}_{-,\plaqindex+1}^\dagger \hat{\beta}_{+,\plaqindex} -i\hat{\beta}_{-,\plaqindex+1}^\dagger \hat{\beta}_{-,\plaqindex} + H.c. \Big].
    \end{split}
    \label{eq:H0_proj}
\end{equation}
Interestingly, this Hamiltonian defines a Creutz ladder
\cite{Creutz1999, Tovmasyan2013}, a well-known all-flat band model sketched in Fig.~\ref{fig:model}(c). 
The destructive interference is determined by the pattern of the hopping phases and accounts for the two flat bands present in the low energy part of the single-particle spectrum of $\hat{H}$.
These results mark an important difference with those of Ref.~\cite{DiLiberto2023}, where $\pi$-flux plaquettes are coupled in a dispersive way in a bosonic version of the BBH model \cite{Benalcazar2017}.
In that case, there are no couplings between $+$ and $-$ orbitals in neighbouring plaquettes, and the projected theory reduces to two independent chains with real hopping strengths.
This allows quasi-condensation and the establishment of long-range chiral order, corresponding to all plaquettes displaying the same chiral polarization.
The presence of flat bands completely changes the picture, as it will be discussed below.

\subsection{Local symmetry}
\label{subsec:symmetry}

All-flat band systems are characterized by the presence of local symmetries \cite{Rontgen2018,Tovmasyan2018}. 
This follows from the fact that the localized Wannier orbitals obtained from a flat band are also energy eigenstates, and their number operator then commutes with the single-particle Hamiltonian. When density-density interactions are taken into account,  it is the parity of the number of particles localized on a  Wannier site (summing over all bands) which is conserved. This argument is expanded in Sec.~\ref{subsec:wannier}.

In this paragraph, we report such local symmetries both in the full and projected models. These can be easily guessed 
(the idea is that swapping $B$ and $C$ sites is a graph automorphism, but one has to correct for the minus sign in the hopping term)
and verified by inspection of the Hamiltonian, or derived using the approach of Ref.~\cite{Tovmasyan2018}.
In the case of the array of $\pi$-flux plaquettes, the set of local symmetries can be built from the operators $\hat{\mathcal{U}}_{\plaqindex, \plaqindex + 1} $, which have a nontrivial action over two consecutive plaquettes,
defined by
\begin{equation}
    \begin{split}
        & \hat{\mathcal{U}}_{\plaqindex, \plaqindex + 1} ^\dagger \, \hat{a}_\plaqindex \,\hat{\mathcal{U}}_{\plaqindex, \plaqindex + 1}  = \hat{a}_\plaqindex, \\
        & \hat{\mathcal{U}}_{\plaqindex, \plaqindex + 1} ^\dagger \, \hat{b}_\plaqindex \, \hat{\mathcal{U}}_{\plaqindex, \plaqindex + 1}  = - \hat{c}_\plaqindex, \\
        & \hat{\mathcal{U}}_{\plaqindex, \plaqindex + 1} ^\dagger \, \hat{c}_\plaqindex \, \hat{\mathcal{U}}_{\plaqindex, \plaqindex + 1}  = - \hat{b}_\plaqindex, \\
        & \hat{\mathcal{U}}_{\plaqindex, \plaqindex + 1} ^\dagger \, \hat{d}_\plaqindex \, \hat{\mathcal{U}}_{\plaqindex, \plaqindex + 1}  = - \hat{d}_\plaqindex, \\
        & \\
        &
        \hat{\mathcal{U}}_{\plaqindex, \plaqindex + 1} ^\dagger \, \hat{a}_{\plaqindex+1} \, \hat{\mathcal{U}}_{\plaqindex, \plaqindex + 1}  = - \hat{a}_{\plaqindex+1}, \\
        & \hat{\mathcal{U}}_{\plaqindex, \plaqindex + 1} ^\dagger \, \hat{b}_{\plaqindex+1} \, \hat{\mathcal{U}}_{\plaqindex, \plaqindex + 1}  = \hat{c}_{\plaqindex+1}, \\
        & \hat{\mathcal{U}}_{\plaqindex, \plaqindex + 1} ^\dagger \, \hat{c}_{\plaqindex+1} \, \hat{\mathcal{U}}_{\plaqindex, \plaqindex + 1}  = \hat{b}_{\plaqindex+1}, \\
        & \hat{\mathcal{U}}_{\plaqindex, \plaqindex + 1} ^\dagger \, \hat{d}_{\plaqindex+1} \, \hat{\mathcal{U}}_{\plaqindex, \plaqindex + 1}  = \hat{d}_{\plaqindex+1};
        \label{eq:full_loc_sym}
    \end{split}
\end{equation}

for any other plaquette, one has $\hat{\mathcal{U}}_{\plaqindex, \plaqindex + 1} ^\dagger \, \hat{\alpha}_{\sigma, k} \, \hat{\mathcal{U}}_{\plaqindex, \plaqindex + 1}  = \hat{\alpha}_{\sigma, k} \,\, \, \forall k \neq \plaqindex , \plaqindex +1$.
These local transformations $\hat{\mathcal{U}}_{\plaqindex, \plaqindex + 1} $ flip the loop currents of the $\plaqindex $-th and $\plaqindex +1$-th plaquettes, namely
\begin{equation}
    \begin{split}
        & \hat{\mathcal{U}}_{\plaqindex, \plaqindex + 1} ^\dagger \, \hat{\mathcal{L}}_{z,\plaqindex} \,\hat{\mathcal{U}}_{\plaqindex, \plaqindex + 1}  = - \hat{\mathcal{L}}_{z,\plaqindex},  \\
        & \hat{\mathcal{U}}_{\plaqindex, \plaqindex + 1} ^\dagger \, \hat{\mathcal{L}}_{z,\plaqindex+1} \, \hat{\mathcal{U}}_{\plaqindex, \plaqindex + 1}  = -\hat{\mathcal{L}}_{z,\plaqindex+1}.
        \label{eq:full_loc_sym}
    \end{split}
\end{equation}
One can easily verify that these operators commute with both the hopping and  the interaction terms of $\hat{H}$, and thus $[\hat{\mathcal{U}}_{\plaqindex, \plaqindex + 1} ,\hat H] = 0$. 
Moreover, notice that, since they swap pairs, the $\hat{\mathcal{U}}$'s are mutually commuting and square to the identity. 
The $\hat{\mathcal{U}}$'s can also be concatenated to obtain the symmetry transformation
$\hat{\mathcal{U}}_{\plaqindex,\plaqindex+r} = \hat{\mathcal{U}}_{\plaqindex, \plaqindex + 1}  \cdot \hat{\mathcal{U}}_{\plaqindex+1,\plaqindex+2} \cdot ... \cdot \hat{\mathcal{U}}_{\plaqindex+r-1,\plaqindex+r}$,
which flips the chirality of only the $\plaqindex$-th and $\plaqindex+r$-th plaquettes.

These local symmetries are inherited by the projected model, acting as follows:

\begin{equation}
    \begin{split}
        & \hat{\mathcal{U}}_{\plaqindex, \plaqindex + 1} ^\dagger \, \hat{\beta}_{-,\plaqindex} \,\hat{\mathcal{U}}_{\plaqindex, \plaqindex + 1}  = -i\hat{\beta}_{+,\plaqindex}, \\
        & \hat{\mathcal{U}}_{\plaqindex, \plaqindex + 1} ^\dagger \, \hat{\beta}_{+,\plaqindex} \, \hat{\mathcal{U}}_{\plaqindex, \plaqindex + 1}  = i \hat{\beta}_{-,\plaqindex}, \\
        & \hat{\mathcal{U}}_{\plaqindex, \plaqindex + 1} ^\dagger \, \hat{\beta}_{-,\plaqindex+1} \, \hat{\mathcal{U}}_{\plaqindex, \plaqindex + 1}  = i \hat{\beta}_{+,\plaqindex+1}, \\
        & \hat{\mathcal{U}}_{\plaqindex, \plaqindex + 1} ^\dagger \, \hat{\beta}_{+,\plaqindex+1} \, \hat{\mathcal{U}}_{\plaqindex, \plaqindex + 1}  = -i \hat{\beta}_{-,\plaqindex+1}, 
        \label{eq:loc_sym}
    \end{split}
\end{equation}
with  $\hat{\mathcal{U}}_{\plaqindex, \plaqindex + 1} ^\dagger \, \hat{\beta}_{\pm k} \, \hat{\mathcal{U}}_{\plaqindex, \plaqindex + 1}  = \hat{\beta}_{\pm k} \,\, \forall k \neq \plaqindex, \plaqindex+1$. 
As a consequence, the angular momentum of the two involved  
plaquettes is flipped

\begin{equation}
    \begin{split}
        & \hat{\mathcal{U}}_{\plaqindex, \plaqindex + 1} ^\dagger \, \hat{L}_{z,\plaqindex} \,\hat{\mathcal{U}}_{\plaqindex, \plaqindex + 1}  = - \hat{L}_{z,\plaqindex},  \\
        & \hat{\mathcal{U}}_{\plaqindex, \plaqindex + 1} ^\dagger \, \hat{L}_{z,\plaqindex+1} \, \hat{\mathcal{U}}_{\plaqindex, \plaqindex + 1}  = -\hat{L}_{z,\plaqindex+1}.
        \label{eq:full_loc_sym}
    \end{split}
\end{equation}

The existence of these set of local symmetries has a dramatic impact on the physics of the system, as it will be explained in the rest of the paper.

\section{Absence of long range Chiral  order}
\label{sec:chiral_orbital_order}

It was demonstrated in Ref.~\cite{DiLiberto2023} that an array of $\pi$-flux plaquettes 
connected in a dispersive way supports long-range chiral order of the Ising type, resulting in 
a ground state with
the loop-currents of each plaquette pointing in the same direction.
In contrast, in our system with Hamiltonian (\ref{eq:H}),  the flatness of the bands and the closely related local symmetries prevent the establishment of such chiral ordering. 
This is enforced by a symmetry argument, which for quantum gauge theories takes the name of Elitzur's theorem, as we will review below.
Furthermore, we will use perturbation theory to derive an effective low-energy spin Hamiltonian, which confirms the absence of chiral order.
Finally, we discuss the ramifications of the local symmetry in the limit of infinite density, where the mean-field theory applies at the level of individual sites, circumventing Elitzur's theorem.

\subsection{Elitzur's theorem}
\label{subsec:Elitzur}

Elitzur's theorem was formulated in the context of
gauge theories~\cite{Elitzur1975,Fradkin2021}. 
It states that in a gauge theory, namely a theory with local constraints or symmetries, the only operators that can have non-vanishing expectation values must be invariant under the local gauge transformations. 
In other terms, local symmetries cannot be spontaneously broken.
This situation is in contrast with the case of systems with global symmetries, where the ground state may spontaneously break this global symmetry.

A more precise statement and a proof of Elitzur's theorem  are reviewed in Appendix \ref{ap:Elitzur}, following \cite{Fradkin2021}.
In short, the essential point is that,
while global symmetry breaking is achieved by sending an external symmetry breaking field to zero after performing the thermodynamic limit, the local symmetry allows to upper bound the effect of the probe field on the expectation value by a local, non-extensive contribution.

As highlighted in Sec. \ref{subsec:symmetry}, the angular momentum operator and its projection are not invariant under the local gauge transformation, and, more specifically, they anti-commute with the local symmetry (when acting on the plaquette of interest). 
Consequently, applying Elitzur's theorem yields 
\begin{equation}
\langle  \hat{\mathcal{L}}_{z,\plaqindex} \rangle   = 0
\ \ \ {\rm and} \ \ \
    \langle  \hat{L}_{z,\plaqindex}  \rangle   = 0.
    \label{eq:zeroL}
\end{equation}
Furthermore, since gauge invariance is preserved, it is straightforward to verify that also the chiral correlators at any two distinct locations vanish:
\begin{equation}
    \langle  \hat{\mathcal{L}}_{z,k} \hat{\mathcal{L}}_{z,\plaqindex} \rangle  = \langle  \hat{\mathcal{L}}_{z,k}
    \hat{\mathcal{U}}_{r-1,r}^\dagger
     \hat{\mathcal{U}}_{r-1,r}
    \hat{\mathcal{L}}_{z,\plaqindex} \rangle 
     = -\langle  \hat{\mathcal{L}}_{z,k} \hat{\mathcal{L}}_{z,\plaqindex} \rangle = 0,
    \label{eq:zeroLL}
\end{equation}
holding  $\forall k \neq \plaqindex $, and  where e.g. $r = \min\{k,\plaqindex\}$, so that $\hat{\mathcal{U}}_{r-1,r}$ anti-commutes with one current operator and commutes with the other one.

\begin{figure}[t]
    \centering
    \includegraphics[scale=0.45]{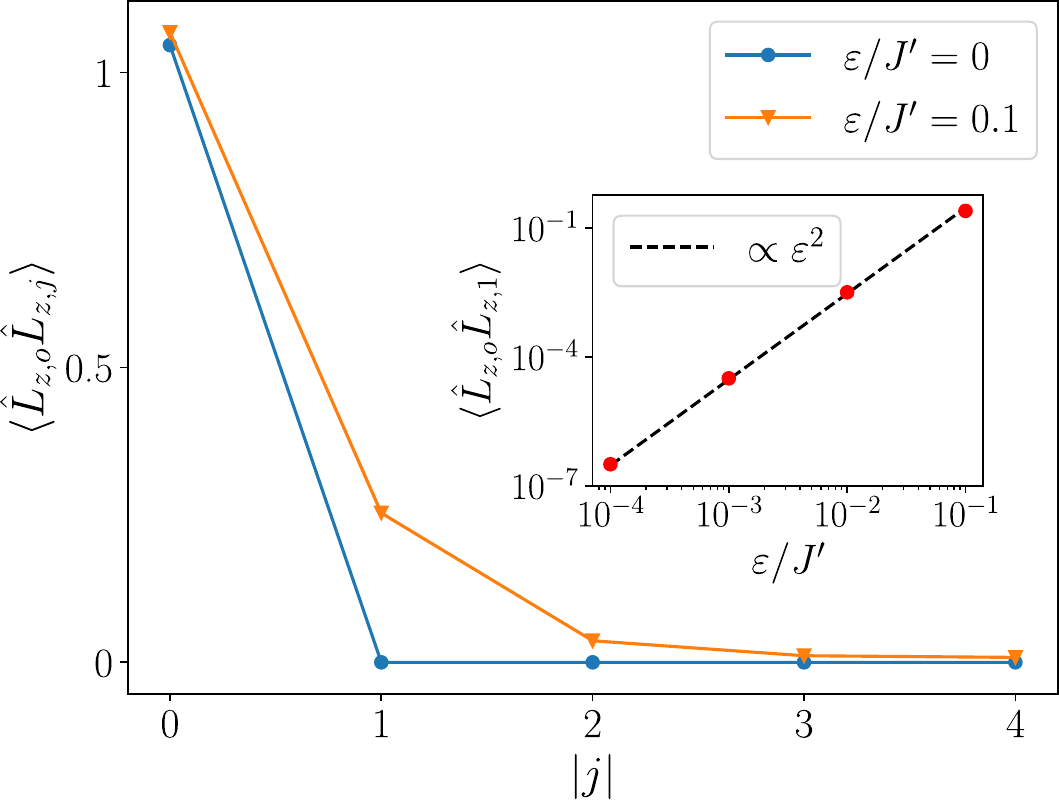}
    \caption{
    The chiral correlator $\langle
    \hat{L}_{z,0} \hat{L}_{z,\plaqindex}
    \rangle$ is shown in blue circles as a function of distance, and it is zero for $\plaqindex \neq 0$ as prescribed by the local symmetry.
     This was obtained via exact diagonalization for the ground state of Hamiltonian (\ref{eq:Hproj_total}) with $L=9$ rungs, $U/J' = 10$.
     The presence of a symmetry breaking term of intensity $\varepsilon$ restores finite correlations (orange triangles).
    The inset
    shows  the increase of the $\filling = 1$ correlator with $\varepsilon$; the black dashes correspond to a $\propto \varepsilon^2$ fit. 
    }
    \label{fig:LzLz}
\end{figure}

The vanishing of chiral order and of the chiral correlators has been verified numerically in Fig. \ref{fig:LzLz}, by computing via exact diagonalization \cite{Weinberg2017, Weinberg2019} the ground state of the projected Creutz ladder model described by Eqs. \eqref{eq:Hint_proj} and \eqref{eq:H0_proj},
with $U/J'=10$ and for $\filling=8$ particles in 8 rungs. 
In the numerics,
we typically truncate the local Hilbert space to a maximum of 4 bosons per site; the validity of this approximation was benchmarked using larger occupations or the full local Hilbert space in smaller systems.

In spite of the sizable chiral fluctuations  
$\langle  \hat{L}_{z,\plaqindex}^2  \rangle$
within a given plaquette, Fig. \ref{fig:LzLz} shows that
the chiral correlator at different sites is  zero (blue circles), up to  machine precision.
As a proof of concept, we have perturbed our model by introducing 
the extra kinetic term $- \varepsilon \sum_\plaqindex (\hat{\beta}_{+,\plaqindex+1}^\dagger \hat{\beta}_{+,\plaqindex} + \hat{\beta}_{-,\plaqindex+1}^\dagger \hat{\beta}_{-,\plaqindex} + \text{h.c.})$ in the Hamiltonian;
we mention that in the unprojected model this can achieved by introducing $BB$ and $CC$ links between adjacent plaquettes.
This perturbation explicitly breaks the local symmetry and results in some dispersion in the Bloch bands of the underlying single-particle spectrum.
As a consequence, the chiral correlator 
$\langle  \hat{L}_{z,k} \hat{L}_{z,\plaqindex} \rangle$
becomes nonzero (orange triangles).
The effect is quadratic in $\varepsilon$,
as visible in the inset.

\subsection{Strongly interacting regime}
\label{subsec:strong_int1}

Above we introduced a general symmetry argument for the absence of long-range chiral order. We now investigate a regime which is particularly convenient for analytical treatment, namely the strongly interacting $U \gg J'$ limit at the integer filling of $\filling$ particles per plaquette. We still require  $U \ll J$ and we perform our analysis in the Creutz ladder projected model.
At zero order of perturbation theory one has $J'=0$, and $2^L$ degenerate ground states of the form 
 $| \Vec{\tau} \rangle 
 = \bigotimes_\plaqindex 
 | \tau_\plaqindex \rangle_\plaqindex
 $, where $ \tau_\plaqindex \in \{+,-\}$
 and
$|\pm \rangle_\plaqindex =\frac{1}{\sqrt{\filling!}} (\beta_{\pm, \plaqindex}^\dagger)^\filling |0\rangle$
as defined above, with $\filling$ the number of bosons per rung.
Perturbation theory leads to processes mixing these states. 
It is convenient to treat this low-energy manifold as a collection of 1/2-spins. Below we discuss separately the $\filling=1$ and $\filling=2$ fillings.

\subsubsection{Filling $\filling=1$}

At this filling, the low energy manifold is 
spanned by the eigenstates
$|\!\!\uparrow\rangle_\plaqindex = \hat{\beta}^\dagger_{+,\plaqindex} |0\rangle_\plaqindex$ and $|\!\!\downarrow\rangle_\plaqindex = \hat{\beta}^\dagger_{-,\plaqindex} |0\rangle_\plaqindex$
of the $\hat{s}^Z_{\plaqindex}$
operator at each rung.
The spin operators are expressed in terms of the Creutz bosonic fields as 
\begin{equation}
\begin{split}
        & \hat{s}^Z_{\plaqindex} = \frac{1}{2}(\hat{\beta}^\dagger_{+,\plaqindex}\hat{\beta}_{+,\plaqindex} - \hat{\beta}^\dagger_{-,\plaqindex}\hat{\beta}_{-,\plaqindex}), \\
        & \hat{s}^X_\plaqindex = \frac{1}{2}(\hat{\beta}^\dagger_{+,\plaqindex}\hat{\beta}_{-,\plaqindex} + \hat{\beta}^\dagger_{-,\plaqindex}\hat{\beta}_{+,\plaqindex}), \\
        & \hat{s}^Y_\plaqindex = \frac{i}{2}(\hat{\beta}^\dagger_{+,\plaqindex}\hat{\beta}_{-,\plaqindex} - \hat{\beta}^\dagger_{-,\plaqindex}\hat{\beta}_{+,\plaqindex}).
\end{split}
\label{eq:1part_spin_op}
\end{equation}
Furthermore, the local symmetry transformations $\hat{\mathcal{U}}_{\plaqindex, \plaqindex + 1} $
anti-commutes with $\hat{s}^Z_{\plaqindex},\hat{s}^X_{\plaqindex},\hat{s}^Z_{\plaqindex+1},\hat{s}^X_{\plaqindex+1}$
and commutes with $\hat{s}^Y_{\plaqindex},\hat{s}^Y_{\plaqindex+1}$, which means that it is represented by $\hat{\mathcal{U}}_{\plaqindex, \plaqindex + 1} =\hat{s}^Y_{\plaqindex} \hat{s}^Y_{\plaqindex+1}$.

Consistently with the presence of the local symmetry, an explicit calculation of the second-order perturbation theory yields  the low-energy spin Hamiltonian
\begin{equation}
    \hat{H}^{\text{eff}}_{\filling=1} =  \frac{J'^2}{4U}\sum_\plaqindex \hat{s}^Y_\plaqindex \hat{s}^Y_{\plaqindex+1}.
    \label{eq:spin_mapp1}
\end{equation}
This corresponds to an anti-ferromagnetic Ising Hamiltonian along the $Y$ direction.
For a ladder with an even number of rungs $L$, the anti-ferromagnetic coupling results in a two-fold ground state degeneracy; for odd $L$ and periodic boundary conditions, instead, the degeneracy is $L$, because a domain wall in the spin-antiferromagnet can be placed in any location of the ladder. This expectation is confirmed by the degeneracies computed numerically, as shown by the blue circles in Fig.~\ref{fig:degeneracy}.
Since the chiral operator $\hat{L}_z$ is given by $\hat{s}^Z$ in the low-energy manifold, no chiral order exists in any of these ground states.

\subsubsection{Filling $\filling = 2$}

\begin{figure}[t]
    \centering
    \includegraphics[scale=0.45]{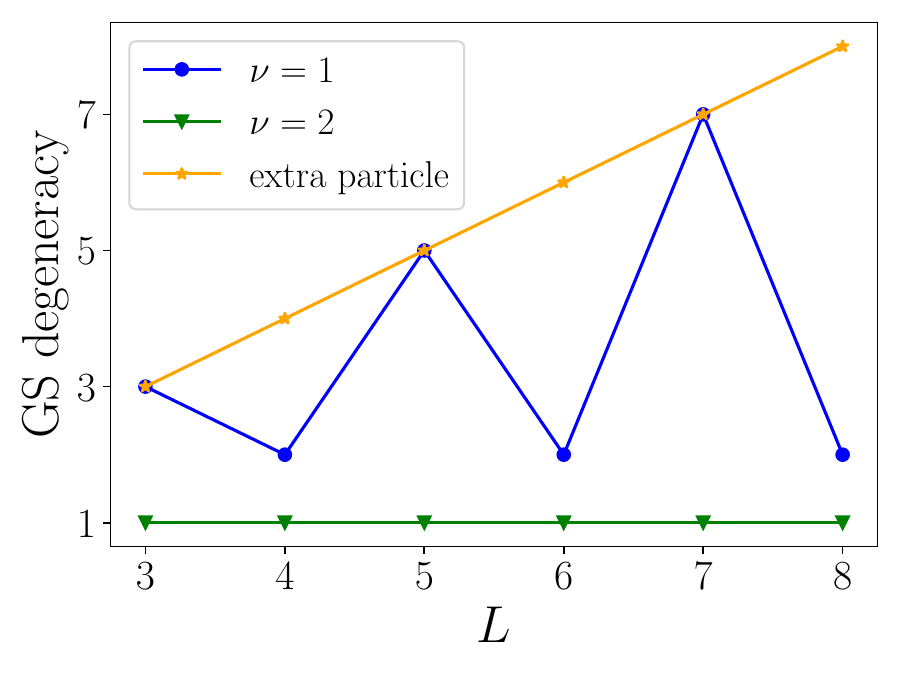}
    \caption{The ground state degeneracy of the projected model Eq. (\ref{eq:Hproj_total}) is obtained via exact diagonalization for $U/J'=10$ and plotted as a function of the number of rungs $L$.
    Blue circles correspond to a filling of $\filling = 1$ particles per rung, and green triangles to $\filling = 2$. These results match the expectations for the effective spin models derived in the $U/J' \gg 1$ limit, i.e. Hamiltonians (\ref{eq:spin_mapp1}) and (\ref{eq:spin_mapp2}), respectively.   
    Orange stars correspond to the case of  $\filling L+1$ bosons. Then, for both $\filling = 1$ and $\filling = 2$, the ground state degeneracy is equal to  the system size and is consistent with the self-pinning behavior of the extra particle.}
    \label{fig:degeneracy}
\end{figure}

In the case of two particles per rung,  
the 1/2 spins  are given by the mapping 
$|\!\!\uparrow\rangle_\plaqindex = \frac{1}{\sqrt{2}}(\hat{\beta}^\dagger_{+,\plaqindex})^2 |0\rangle_\plaqindex$
and ~ ~ ~ ~ ~ ~ ~ ~ ~ ~ ~ ~ ~ ~ ~ ~ ~
$|\!\!\downarrow\rangle_\plaqindex = \frac{1}{\sqrt{2}}(\hat{\beta}^\dagger_{-,\plaqindex})^2 |0\rangle_\plaqindex$
, and
the spin operators describing the low-energy manifold
are expressed by
\begin{equation}
\begin{split}
        & \hat{S}^Z_\plaqindex = \frac{1}{4}(\hat{\beta}^\dagger_{+,\plaqindex}\hat{\beta}^\dagger_{+,\plaqindex}\hat{\beta}_{+,\plaqindex}\hat{\beta}_{+,\plaqindex} - \hat{\beta}^\dagger_{-,\plaqindex}\hat{\beta}^\dagger_{-,\plaqindex}\hat{\beta}_{-,i}\hat{\beta}_{-,\plaqindex}), \\
        & \hat{S}^X_\plaqindex = \frac{1}{4}(\hat{\beta}^\dagger_{+,\plaqindex}\hat{\beta}^\dagger_{+,\plaqindex}\hat{\beta}_{-,\plaqindex}\hat{\beta}_{-,\plaqindex} + \hat{\beta}^\dagger_{-,\plaqindex}\hat{\beta}^\dagger_{-,\plaqindex}\hat{\beta}_{+,\plaqindex}\hat{\beta}_{+,\plaqindex}), \\
        & \hat{S}^Y_\plaqindex = \frac{i}{4}(\hat{\beta}^\dagger_{+,\plaqindex}\hat{\beta}^\dagger_{+,\plaqindex}\hat{\beta}_{-,\plaqindex}\hat{\beta}_{-,\plaqindex} - \hat{\beta}^\dagger_{-,\plaqindex}\hat{\beta}^\dagger_{-,\plaqindex}\hat{\beta}_{+,\plaqindex}\hat{\beta}_{+,\plaqindex}).
\end{split}
\label{eq:2part_spin_op}
\end{equation}
In this case, the action of the local symmetry is given by $\hat{\mathcal{U}}_{\plaqindex, \plaqindex + 1} =\hat{S}^X_{\plaqindex} \hat{S}^X_{\plaqindex+1}$,
and fourth order perturbation theory yields the effective spin Hamiltonian
\begin{equation}
    \hat{H}^{\text{eff}}_{\filling=2} = - \frac{J'^4}{U^3}\sum_\plaqindex \bigg( h \hat{S}^X_\plaqindex + \frac{1}{4} \hat{S}^X_\plaqindex  \hat{S}^X_{\plaqindex+1} \bigg),
    \label{eq:spin_mapp2}
\end{equation} 
with $h = \frac{365697}{42930} \simeq 8.52$.
We remark the presence of the (strong) external field in the $X$ direction. 
This corresponds to the flipping of a boson pair across a rung, mediated by pair hopping through adjacent sites.
On the contrary, an external field is not present in $\filling = 1$ effective Hamiltonian (\ref{eq:spin_mapp1}), since the flipping of single bosons is prevented by destructive interference (processes occurring on the right of a rung cancel the ones on the left side).
The external field polarizes the spins along the $X$ direction. In this case, the ground state is unique (as confirmed by the numerics in Fig.~\ref{fig:degeneracy}, green triangles), and once again no chiral order is present along the $Z$ direction, as prescribed by Elitzur's theorem. 

~

\begin{figure}[t]
    \centering
    \includegraphics[scale=0.42]{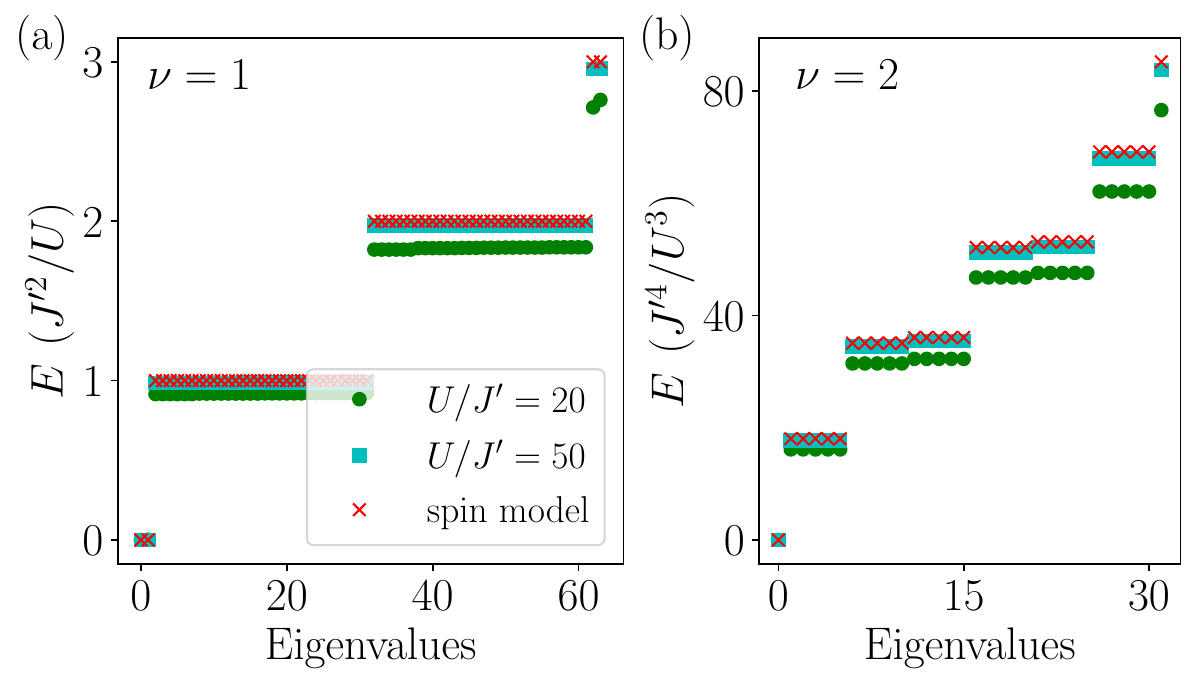}
    \caption{
    The low part of the energy spectrum of the projected Hamiltonian (\ref{eq:Hproj_total}) 
    has been calculated via exact diagonalization and 
    is compared to the spectrum of the spin Hamiltonians obtained from perturbation theory. 
    (a) and (b) panels correspond to
   $ \filling = 1$ and  $\filling = 2$ particles per rungs, respectively, with the spin models given by Eq.~(\ref{eq:spin_mapp1}) and Eq.~(\ref{eq:spin_mapp2}).
    Convergence towards the spin model predictions is observed when increasing $U/J'$, as highlighted by measuring the energies in units of $J'^2/U$ and $J'^4/U^3$. 
   }
    \label{fig:E_spin_mapping}
\end{figure}

Furthermore, for both fillings $\filling = 1,2$ we computed the full many-body spectrum using exact diagonalization for a short ladder. In Fig.~\ref{fig:E_spin_mapping} 
we compare the low energy part of the excitation spectrum with the predictions of the effective spin Hamiltonians of Eq.~(\ref{eq:spin_mapp1}) in panel (a) and of Eq.~(\ref{eq:spin_mapp2}) in panel (b).
Convergence is observed for increasing $U/J'$, providing a numerical validation of the effective spin models.

\subsection{Mean-field regime}
\label{subsec:MF}

\begin{figure*}[t]
    \centering
    \includegraphics[width=\textwidth]{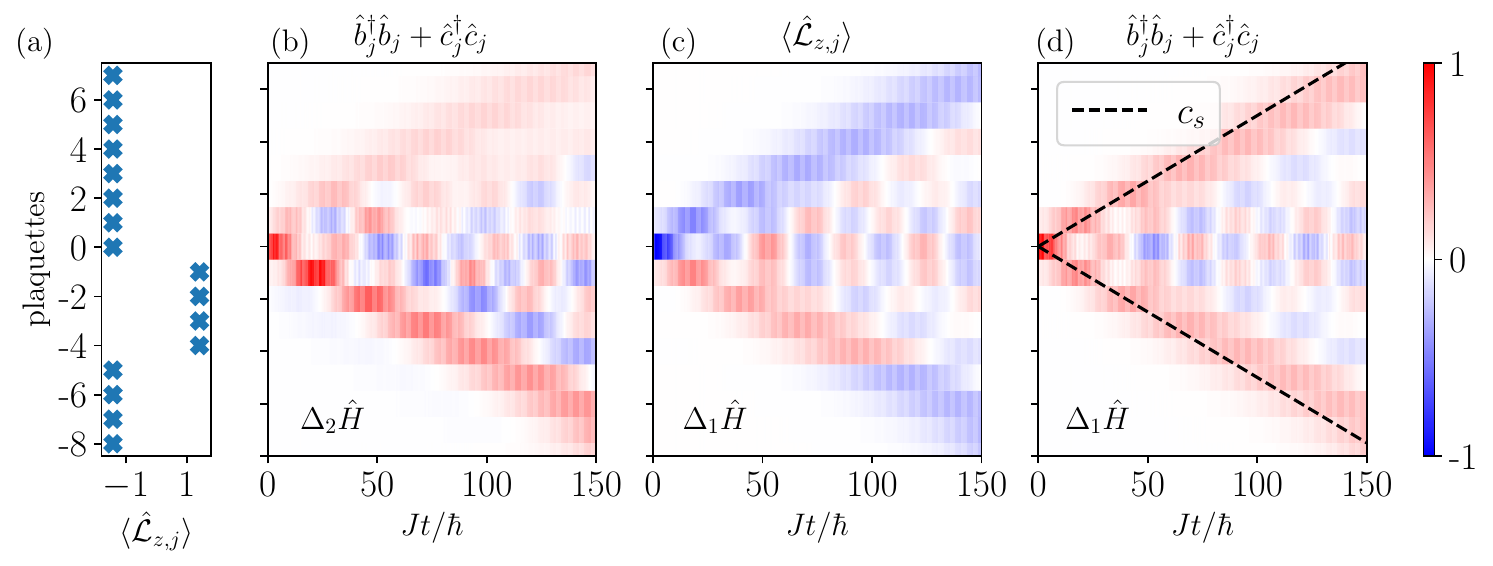}
    \caption{
    Gauge-invariance and mean-field dynamics. (a) The chiral polarization pattern of the mean-field state, which we selected arbitrarily out of $2^L$ possibilities.
    (b) The not gauge invariant  perturbation $\Delta_2 \hat{H}$ is switched off at time $t=0$, and  
    the evolution of the gauge invariant observable $|b_\plaqindex|^2 + |c_\plaqindex|^2$ (or rather its variation with respect to the unperturbed mean-field state) is monitored.
The spatially asymmetric behavior reflects the asymmetry in the selected mean-field state.
(c) Same protocol as in (b), but for the gauge invariant perturbation $\Delta_1 \hat{H}$ and non-invariant observable $\langle \hat{\mathcal{L}_z} \rangle$. The asymmetry persists.
(d) Instead, if both the perturbation and observable are gauge invariant, the results are independent of the initial configuration choice and spatially symmetric. The dashed lines denote the cone predicted from the Bogoliubov speed of sound $c_s$ of the projected model. 
In all panels we used $J' = 0.1J$, $g \equiv U\filling = 0.1J$ and $\Delta = 0.001J$.}
    \label{fig:MF_evol}
\end{figure*}

Elitzur's theorem and the low-energy analysis of the previous sections rule out the onset of chiral order, meaning that $\langle \hat{\mathcal{L}}_{z,\plaqindex} \rangle = 0$, as well as its projected version 
$\langle \hat{L}_{z,\plaqindex} \rangle = 0$ and similarly for the correlators at different sites.
Therefore, a mean-field analysis based on a  chiral order parameter seems inappropriate.
However, when the number of bosons per plaquette $\filling$ is very large and interactions are very weak $U \ll J$, while imposing the condition that $U \filling$ remains finite, 
mean-field behavior and strong chiral fluctuations may persist on finite time scales or in the presence of symmetry breaking perturbations \cite{DiLiberto2023}.
This could be the situation in ultracold gases or in polariton and optical devices~\cite{amelio2024lasing}.

For these reasons, we now investigate the mean-field dynamics in  a Gross-Pitaevskii framework, where bosonic operators are replaced by complex numbers. Since the numerics is cheap within this approach, we work in the unprojected model of $\pi$-flux plaquettes.
It is straightforward to verify that the local symmetry extends to the Gross-Pitaevskii equation. As a consequence, minimization of the energy in the regime $J \gg U \filling, J'$ leads to $2^L$ degenerate mean-field ground states, corresponding to two possible chiral polarizations for each plaquette. 
More details about the mean-field approach are available in the Appendix \ref{ap:MF}.
We remark once again that only in the large $\filling$ limit Elitzur's theorem can be circumvented, 
and these states can be stabilized by small local chiral symmetry perturbations,
while quantum fluctuations lead to their mixing and a paramagnetic response in the $\filling \sim O(1)$ case, as studied in the previous subsections. \\

We observe that, even though gauge invariance is broken by having selected a   
mean-field ground state with some pattern of chiral polarization, it is still possible to obtain gauge invariant information by measuring the correlator of two 
gauge invariant quantities.
More precisely, we start by arbitrarily selecting one mean-field ground state with the polarization pattern shown in Fig.~\ref{fig:MF_evol}(a).
We then consider two kinds of perturbations in the Hamiltonian,
first, the gauge invariant operator 
$\Delta_1 \hat{H} = -\Delta (\hat{b}_{0}^\dagger \hat{b}_{0} + \hat{c}_{0}^\dagger \hat{c}_{0})$, and,  second, the non-invariant quantity 
$\Delta_2 \hat{H} = -\Delta \hat{b}_{0}^\dagger \hat{b}_{0}$, where $\Delta \ll J,J',U \filling$.
Both perturbations are confined to the 0-th plaquette 
in particular, the local symmetry maps $\Delta_2 \hat{H}$
to $
\hat{\mathcal{U}}_{\plaqindex, \plaqindex + 1} ^\dagger
\Delta_2 \hat{H} 
\hat{\mathcal{U}}_{\plaqindex, \plaqindex + 1} 
= 
 -\Delta \hat{c}_{0}^\dagger \hat{c}_{0}
$, where $\plaqindex \in \{ -1, 0\}$. \\
The perturbed ground state is found via imaginary time evolution.
We then switch off the perturbation and perform real time evolution.
Two observables are monitored for each plaquette, the $BC$ density $\hat{b}_{\plaqindex}^\dagger \hat{b}_{\plaqindex} + \hat{c}_{\plaqindex}^\dagger \hat{c}_{\plaqindex}$ and 
the loop-current $\hat{\mathcal{L}}_{z,\plaqindex}$, only the first one being gauge invariant.
When either the perturbation or the observable is not gauge invariant, the initially selected mean-field state matters, as visible in the spatially asymmetric signal in Fig.~\ref{fig:MF_evol}(b) and (c).
Instead, the dynamics of the gauge invariant observable upon a gauge invariant perturbation does not depend on the choice of the initial chiral configuration, resulting in the symmetric signal in Fig.~\ref{fig:MF_evol}(d) (this has been verified also with other mean-field configurations, not shown).
This illustration suggests how to extract gauge invariant information from a mean-field treatment which explicitly breaks gauge invariance.

\section{Impurity  self-pinning}
\label{sec:pair_superfluidity}

Previous studies of flat band systems demonstrated that Aharanov-Bohm caging inhibits single-particle transport, while pairs can disperse and even give rise to pair superfluidity~\cite{Takayoshi2013,Tovmasyan2013,Tovmasyan2018,Danieli2021}.
In this Section we report on a phenomenon that is in contrast with the physics of pair
superfluidity, which we dub ``impurity self-pinning''.
 This expression refers to the fact that, 
when an extra boson is added on top of an integer filling ground-state, the density of the system in the presence of the  impurity  remains localized. We stress that this localization is not due to caging nor symmetry, since pair transport is in principle allowed, but arises from interactions and energetics.
Numerical calculations suggest that this self-pinning phenomenon is quite generic in our model, but  it is analyzed below in details in the $U \ll J'$ regime. To this end, we first recall the Wannier orbitals in the Creutz ladder.

\subsection{Wannier basis}
\label{subsec:wannier}

To analyze the weakly interacting $U \ll J'$ regime, it is convenient to work in the Wannier basis.
 For the Creutz ladder, it is easy to derive~\cite{Tovmasyan2013, Takayoshi2013} the Wannier annihilation operators  
\begin{equation}
    \hat{w}_{\gamma,\plaqindex} = \frac{1}{2}(\hat{\beta}_{+,\plaqindex} - i\hat{\beta}_{-,\plaqindex} - (-1)^\gamma i\hat{\beta}_{+,\plaqindex+1} + (-1)^\gamma \hat{\beta}_{-,\plaqindex+1}),
    \label{eq:wannier_operator}
\end{equation}
where  $\gamma = 0,1$
labels the first and second energy band, respectively.
The $\plaqindex$-th Wannier orbital 

$w_\gamma(\sigma, \plaqindex) = \langle 
\sigma,\plaqindex |
\hat{w}_{\gamma,\plaqindex}^\dagger
| 0
\rangle
$
is localized on the $\plaqindex$-th and $\plaqindex+1$-th rung, with the amplitudes indicated in Fig.~\ref{fig:model}(c).
The Wannier states are mutually orthogonal and, in flat band systems, they are eigenstates of the Hamiltonian.
The action of the local symmetry on the Wannier operators reads 
\begin{equation}
         \hat{\mathcal{U}}_{\plaqindex, \plaqindex + 1} ^\dagger \, \hat{w}_{\gamma,k} \,\hat{\mathcal{U}}_{\plaqindex, \plaqindex + 1}  = (-1)^{\delta_{k,\plaqindex}} \ \hat{w}_{\gamma,k}.
\end{equation}
As pointed out in Ref. \cite{Tovmasyan2018}, the local symmetry is nothing but the parity operator 
\begin{equation}
    \hat{\mathcal{U}}_{\plaqindex, \plaqindex + 1}  = \exp \Bigg( i\pi \sum_{\gamma = 0,1 } \hat{w}_{\gamma,\plaqindex}^\dagger \hat{w}_{\gamma,\plaqindex} \Bigg),
    \label{eq:symm_wannier}
\end{equation}
which reflects the parity of the 
number of particles in the Wannier orbitals at a given location, keeping into account all bands. In this representation, it is particularly clear that single particle transport flips the parity and is forbidden by symmetry.

Furthermore, as a consequence of Eq.~(\ref{eq:symm_wannier}) and of Elitzur's theorem, 
we expect single-particle coherences $\langle \hat{w}_{\gamma,k}^\dagger \hat{w}_{\gamma',\plaqindex} \rangle$ to be zero
at different sites $k \neq \plaqindex$;
the proof is perfectly analogous to the one in Eq.~\eqref{eq:zeroLL}.
This rules out conventional condensation or quasi-condensation.
However, in general the pair correlator 
$\langle \hat{w}_{\gamma,k}^\dagger \hat{w}_{\gamma',k}^\dagger \hat{w}_{\gamma'',\plaqindex} \hat{w}_{\gamma''',\plaqindex} \rangle$
is non-zero,
and indeed pair-superfluidity has been predicted~\cite{Tovmasyan2013,Takayoshi2013,Tovmasyan2018}
in the Creutz ladder with contact repulsive interactions.

\subsection{Impurity self-pinning  for $J' \gg U$}
\label{subsec:weak_int}

For weak interactions $J' \gg U$, the dynamics of the system can be projected into the lowest single-particle band, described by the Wannier operators $\hat{w}_{0,\plaqindex}$. Dropping from now on the subscript and neglecting a constant energy term $-J'/2$ per particle, 
we get the single-band Hamiltonian 
\begin{equation}
    \hat{H}^{\rm W} = U \sum_\plaqindex \Big( \frac{3}{4} \hat{\rho}_\plaqindex (\hat{\rho}_\plaqindex - 1) + \frac{1}{2} \hat{\rho}_\plaqindex \hat{\rho}_{\plaqindex+1} + \frac{1}{8} (\hat{w}^{\dagger 2}_{\plaqindex} \hat{w}_{\plaqindex+1}^2 + \text{c.c.}) \Big),
    \label{eq:H_wannier}
\end{equation}
where we introduced the Wannier density $\hat{\rho}_{\plaqindex} \equiv \hat{w}_{\plaqindex}^\dagger \hat{w}_{\plaqindex}$. The first term corresponds to a standard onsite interaction, while the second term describes nearest-neighbor repulsion. Finally, the last term is responsible for  pair hopping processes between two successive sites. Single-particle hopping is instead forbidden by symmetry,
since 
$
\hat{\mathcal{U}}_{\plaqindex-1,\plaqindex}^\dagger
\hat{w}^{\dagger}_{\plaqindex} \hat{w}_{\plaqindex+1}
\hat{\mathcal{U}}_{\plaqindex-1,\plaqindex}
=-\hat{w}^{\dagger}_{\plaqindex} \hat{w}_{\plaqindex+1}
$.

We remark the subtle but essential difference of the projected Hamiltonian (\ref{eq:H_wannier}) with respect to the effective 
Hamiltonians obtained in previous studies of the Creutz ladder with contact interactions~\cite{Tovmasyan2013,Takayoshi2013,Tovmasyan2018}. In these works a very similar low-energy Hamiltonian is obtained,

$
\hat{H}^{\rm W}_{\rm std} = U \sum_\plaqindex \Big( \frac{1}{4} \hat{\rho}_\plaqindex (\hat{\rho}_\plaqindex - 1) + \frac{1}{2} \hat{\rho}_\plaqindex \hat{\rho}_{\plaqindex+1} + \frac{1}{8} (\hat{w}^{\dagger 2}_{\plaqindex} \hat{w}_{\plaqindex+1}^2 + \text{c.c.}) \Big)
$, identical to  $\hat{H}^{\rm W}$ but with the coefficient $1/4$ instead of $3/4$ in the first term. 
This difference is due to the  presence of the
angular-momentum interaction 
in Eq.~(\ref{eq:Hint_proj}), which 
turns out to
unfavor pair formation.
As a consequence, at filling $\filling = 1$ the ground-state of 
$\hat{H}^{\rm W}$
features
\textit{exactly}
one particle per Wannier orbital, i.e. $|GS\rangle = \prod_\plaqindex \hat{w}^\dagger_\plaqindex |0\rangle$. 
Moreover, at fillings $\nu < 1$, the ground state is degenerate and can be expressed as a  tensor product state with 0 or 1 particles per site.
Instead, in the standard case $\hat{H}^{\rm W}_{\rm std}$  with no angular momentum interactions, one would expect at $\nu = 1$ the pair charge density wave
$|GS\rangle \propto \prod_{n \ {\rm odd}} (\hat{w}^\dagger_\plaqindex)^2 |0\rangle$, and pair superfluidity emerges for fillings $0.70 \leq \filling < 1$~\cite{Takayoshi2013}.
 It is worth mentioning Ref.~\cite{Lühmann2016}, where the  Wannier onsite repulsion is varied at filling $\filling = 4$, encompassing, in the order,  pair superfluid,  pair Mott and  Mott insulator phases for increasing $\hat{\rho}_\plaqindex^2$ coupling.

\begin{figure}[t]
    \centering
    \includegraphics[scale=0.37]{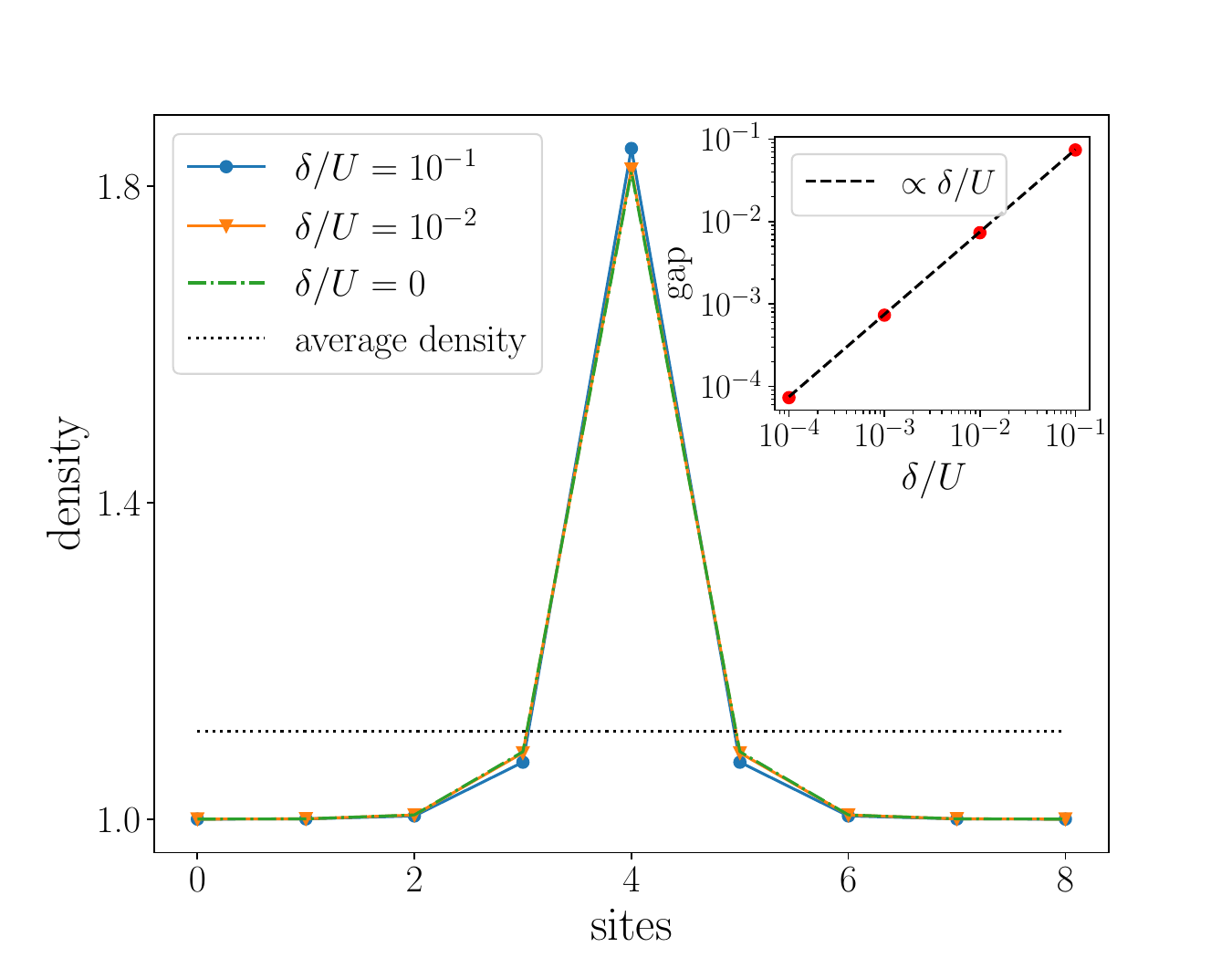}
    \caption{
    Density of the ground state of Hamiltonian \eqref{eq:H_wannier}
    in the presence of $L+1$ bosons in $L=9$ rungs, as obtained from exact diagonalization.
    A small pinning potential of depth $\delta$ is introduced to break translational symmetry and resolve the degeneracy of the ground state manifold; the density converges to a peak for vanishing $\delta$, instead of melting to the average density value (horizontal dotted line) as expected in a dispersive system. The $\delta /U = 0$ case (green dotted line) is computed by diagonalizing $\hat{H}^\psi$ from Eq. \eqref{eq:doublet&hole}.
    Inset:  the energy gap between the ground state and the first excited state increases linearly in $\delta$.
    These observations demonstrate that  ``self-pinned states'' exist in the ground-state manifold already at $\delta/U=0$.}
    \label{fig:pinning_pot_wannier}
\end{figure}

We focus here on an ``impurity self-pinning'' phenomenon characteristic of $\hat{H}^{\rm W}$.
Starting from an integer filling, we  add an extra boson to the system and find, via exact diagonalization, that the ground state manifold is $L$-fold degenerate (see orange stars in Fig.~\ref{fig:degeneracy}), and consists of states in which the excess density  can be localized with zero energy cost. 
This is demonstrated by adding a small pinning potential to the Hamiltonian, and computing the ground state density. The pinning potential removes the ground state degeneracy and localizes the excess particle at the potential location.
When the pinning intensity $\delta/U$ is decreased, the ground-state density converges to an exponentially localized peak, as shown in Fig.~\ref{fig:pinning_pot_wannier}. Moreover, the energy gap between the ground and first excited states scales linearly in $\delta/U$.
If the impurity were dispersive, the 
density peak would disappear for vanishing $\delta$ and the gap would scale  quadratically~\footnote{
In an infinite system, this can be for instance seen from the textbook calculation of a particle in a Dirac delta attractive well
in one dimension~\cite{Griffiths2018}; in this case, the bound state decays exponentially with a characteristic length proportional to the potential strength.
In a finite system, instead, a dispersive particle has a discrete spectrum with a non-degenerate ground state at zero momentum. First order perturbation theory then applies, and a small delta potential would then induce a linear energy shift scaling as the inverse system size, but would not lead to any sizable localization.
}.

\begin{figure}[t]
    \centering
    \includegraphics[scale=0.42]{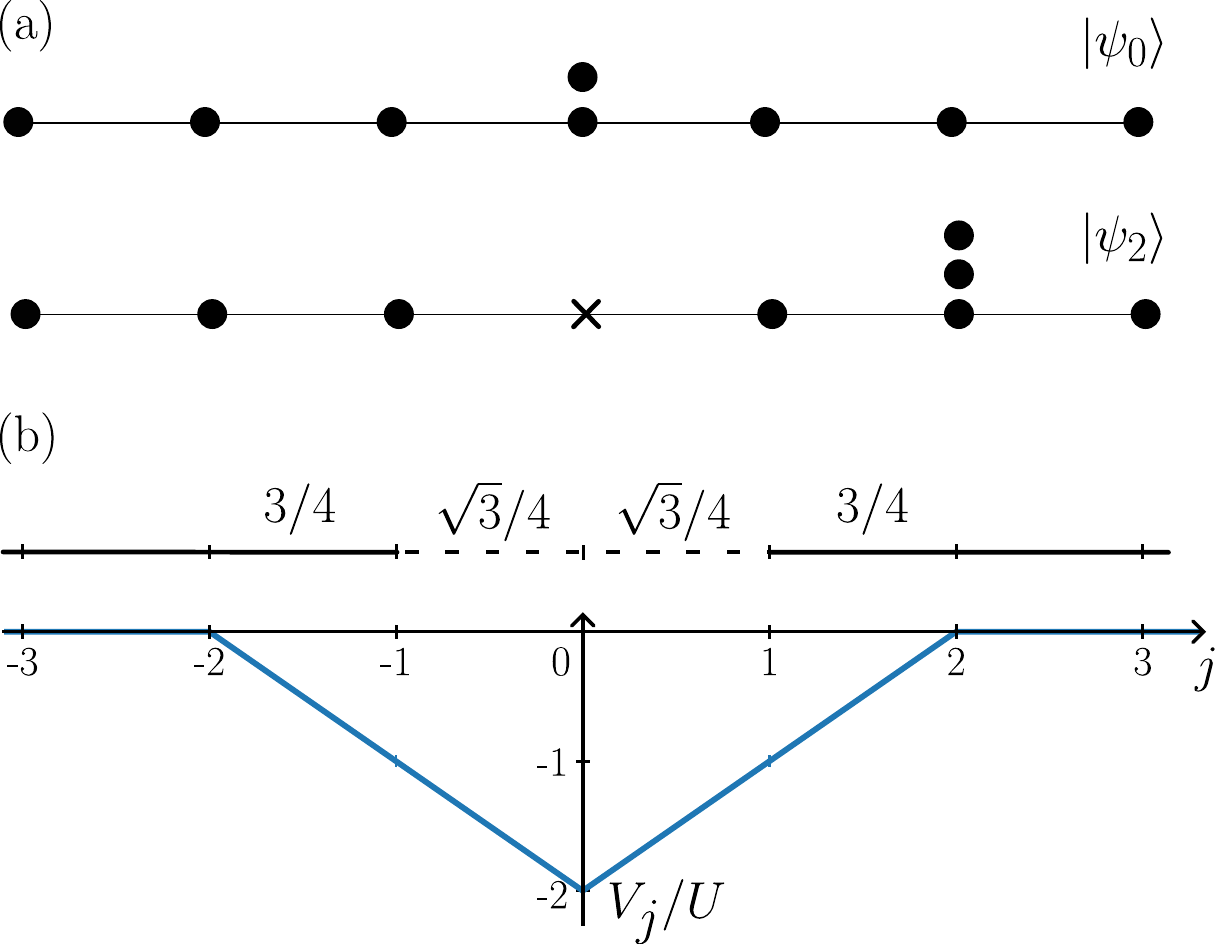}
    \caption{Schematic representation of the subspace 
    $\{ |\psi_\plaqindex \rangle \}_{\plaqindex \in \mathbb{Z}}$  and the associated Hamiltonian $H^\psi$. (a) The two states $\ket{\psi_0}$ and $\ket{\psi_2}$ are shown, with the number of black dots indicating the bosonic occupations. (b) The Hamiltonian $H^\psi$ in Eq. \eqref{eq:doublet&hole}
    effectively describes a particle hopping on a chain  and subject to the attractive potential $V_\plaqindex$ centered on site $j = 0$.
    The links involving $j = 0$ have a different hopping strength, highlighted by the dashed lines.
    }
    \label{fig:model_wannier}
\end{figure}

These observations define the self-pinning of the excess particle.
We remark that such phenomenology is \textit{not} a direct consequence of Aharanov-Bohm caging, since pair hopping is possible in principle, and would delocalize the excess density.
While we numerically observed a similar physical behavior independently 
of a small $U/J'$ and
at other integer fillings (not shown),
the present regime $U \ll J'$ allows for nearly analytical treatment.
We start by considering the projected Hamiltonian (\ref{eq:H_wannier}) and the filling 1 ground state $|GS\rangle = \prod_\plaqindex \hat{w}^\dagger_\plaqindex |0\rangle$. We then add a boson  and assume an infinitesimal pinning potential at site $\plaqindex = 0$. 
One can verify that the new ground state 
lives in the subspace spanned by the (normalized) states 
\begin{equation}
    |\psi_\plaqindex\rangle = \frac{1}{\sqrt{2+4\delta_{\plaqindex,0}}} (\hat{w}^\dagger_\plaqindex)^2 \prod_{k \neq 0} \hat{w}^\dagger_k |0\rangle.
    \label{eq:psi_subspace}
\end{equation}
In other words, for $\plaqindex \neq 0$ the state $|\psi_\plaqindex\rangle$ contains one boson per Wannier site, but  3 bosons in site $j$ and  no bosons in site 0.
Instead,  $|\psi_0\rangle$ contains a doublon at 0. As an illustration, the states 
$|\psi_0 \rangle$
and $|\psi_2 \rangle$
are 
schematically
shown in Fig. \ref{fig:model_wannier}(a). 
The subspace 
$\{ |\psi_\plaqindex \rangle \}_{\plaqindex \in \mathbb{Z}}$
is closed under the action of the Hamiltonian,
since only pair-hopping is allowed, and
 we can write without approximations (we just leave out a constant term)
\begin{equation}
\begin{split}
      \hat{H}^{\psi} 
      \equiv
      \hat{P}_\psi \hat{H}^{W} \hat{P}_\psi
      = & \,
      U \bigg[ \frac{\sqrt{3} - 3}{4} \Big( \ket{\psi_1} \bra{\psi_0} + \ket{\psi_0} \bra{\psi_{-1}} \Big)  \\
      & - \ket{\psi_{\pm 1}} \bra{\psi_{\pm 1}} - \textcolor{red}{3}\ket{\psi_0} \bra{\psi_0} \\
      & + \frac{3}{4}  \sum_\plaqindex |\psi_{\plaqindex+1}\rangle \langle \psi_\plaqindex| \bigg] + H.c. ,
\end{split}
\label{eq:doublet&hole}
\end{equation}
where we introduced the projector $\hat{P}_\psi = \sum_\plaqindex |\psi_\plaqindex\rangle \langle \psi_\plaqindex |$.
This is the  Hamiltonian of a dispersive particle in an attractive potential $V_\plaqindex = \langle \psi_\plaqindex | \hat{H}^{\psi} | \psi_\plaqindex \rangle$, centered  around $\plaqindex = 0$, as shown in Fig. \ref{fig:model_wannier}(b), and supports a bound state. Diagonalizing $\hat{H}^{\psi}$ yields the density profile 
displayed in Fig.~\ref{fig:pinning_pot_wannier} under the label $\delta=0$. This shows that the self-pinning phenomenon 
arises from the binding of an auxiliary particle described by $\hat{H}^{\psi}$, and can be thought of as the bound state of a bosonic pair with a self-generated hole at $\plaqindex = 0$.

Finally, at larger fillings, preliminary studies suggest that the ground state is chracterized by the formation of many self-pinned pairs.

\section{Conclusions}
\label{sec:conclusion}

In conclusion, we have demonstrated the crucial role of local symmetries in preventing the establishment of chiral order in an all-flat band chain of $\pi$-flux plaquettes. 
We have also identified a self-pinning phenomenon that originates from the interplay of many-body interactions and local symmetry. This exquisite many-body localization phenomenon appears to elude the mechanism of Aharonov-Bohm caging.

Our model could be implemented in ultracold atom experiments, where precise control over interactions and the ability to realize complex hopping phases and synthetic gauge fields make it an ideal platform~\cite{Lin2009, Aidelsburger2011, Aidelsburger2013}. The Aharonov-Bohm effect was thus observed with ultracold atoms in a diamond chain in \cite{Li2022}.
In the realm of photonics, artificial magnetic fields are achieved through modulation-assisted tunneling, enabling the engineering of similar all-flat band systems~\cite{Mukherjee2018}.
There are also ongoing efforts to realize the bosonic Creutz ladder using optical waveguides, where synthetic dimensions could play a crucial role~\cite{Zurita2020}.
Another candidate platform is provided by superconducting circuits,
where
chiral ground-state currents of interacting photons were realized in
\cite{Roushan2016}; furthermore, 
a single $\pi$-flux plaquette was recently realized in ~\cite{martinez2023flatband}, where they also demonstrated the pair-hopping-induced delocalization for two interacting photons.

In future theoretical works, it would be interesting to identify tight-binding models featuring non-Abelian local symmetries, which may have some relations with quantum fields studied
in particle physics.
Also, it would be very interesting to investigate the connections between the self-pinning observed here and other many-body localization phenomena driven by interactions~\cite{Abanin2019}, including  confinement of fractons~\cite{boesl2024deconfinement} 
or disorder-free localization~\cite{Brenes2018, Halimeh2022}.
We also point out that our model can be extended to higher dimensions; see Appendix \ref{ap:2D}. The presence of local symmetry, as well as the consequences of Elitzur theorem, are shown to directly transpose to this higher-dimensional context; the existence of the self-localization phenomenon is still to be quantitatively assessed.

Finally, studying thermalization, information spreading  and quantum scars~\cite{Turner2018,moudgalya2020,Moudgalya2022} in this family of models would be very intriguing.

\section*{Acknowledgements}

We thank Felix Palm, Lucila Peralta Gavensky,  and Subir Sachdev for useful discussions.
Work in Brussels is supported by the ERC Grant LATIS, the EOS project CHEQS, the Fonds de la Recherche Scientifique (F.R.S.-FNRS), the Fondation ULB and the Fondation Roi Baudouin. 
Work in Padua is supported by the Italian Ministry of University and Research via the Rita Levi-Montalcini program, and by the European Union via the H2020 Quantum Flagship project PASQuanS2 and the QuantERA project T-NiSQ. Exact diagonalization simulations are done using the Python package QuSpin \cite{Weinberg2017, Weinberg2019}. \\
Data availability—The data that support the findings
of this article and the codes to produce them are openly
available at \cite{Burgher_DATA}.

\appendix

\section{Elitzur's theorem}
\label{ap:Elitzur}

In this appendix, we present a general formulation of Elitzur's theorem in a gauge theory based on the proof given in \cite{Fradkin2021}. We then apply it to our model and its projection to show that no chiral order can be established in presence of local symmetry.

More specifically, it is assumed that a compact symmetry group exists, whose elements have a nontrivial action only on a finite number of degrees of freedom. One then considers a local operator 
and its average under the action of the group, 
i.e.
the average on the states  obtained by applying all the transformations of the group.
If this average is zero, then the 
expectation value of such an operator must be zero.
Let's call 
$\hat{\mathcal{L}}_0$ the local observable, and for simplicity let's consider a local symmetry group containing only the identity and the operator $\hat{\mathcal{U}}$ which anti-commutes with $\hat{\mathcal{L}}_0$,
i.e.
$\hat{\mathcal{U}}^\dagger \mathcal{L}_0 \hat{\mathcal{U}} = - \hat{\mathcal{L}}_0$.

Inspired by the fact that global symmetry breaking is achieved by sending an external symmetry breaking field to zero after performing the thermodynamic limit, a similar approach can be adopted in the case of a gauge theory.
More precisely, one needs to compute
\begin{equation}
   \langle \hat{\mathcal{L}}_0 \rangle
   =
   \frac{1}{Z} {\rm Tr}
   \left[
   \hat{\mathcal{L}}_0 e^{-\beta \hat{H} + h \sum_j \hat{\mathcal{L}}_j}
   \right],
\end{equation}
where $\beta$ is the inverse temperature (which we take arbitrarily large here), 
$j$ is a generic spatial index,
$h$ denotes the probe symmetry breaking field,
$Z$ the partition function,
and the Hamiltonian commutes with the local symmetry $[\hat{H}, \hat{\mathcal{U}}]=0$.
Moreover, locality entails that 
$[\hat{\mathcal{L}}_j, \hat{\mathcal{U}}]=0$, with at most a finite number of exceptions $K$ (including the case $j=0$).
The symmetry entails that
$\langle \hat{\mathcal{L}}_0 \rangle
= \langle -\hat{\mathcal{L}}_0
e^{h \sum_j (
\hat{\mathcal{U}}^\dagger
\hat{\mathcal{L}}_j
\hat{\mathcal{U}} - \hat{\mathcal{L}}_j ) }
\rangle$,
yielding the bound
\begin{equation}
    2|\langle \hat{\mathcal{L}}_0 \rangle|
\leq 
|| \hat{\mathcal{L}}_0 ||
\left(
e^{2hK||\hat{\mathcal{L}}_0||} - 1
\right),
\label{eq:Elitzur_bound}
\end{equation}
which holds for any finite $h,L$ and is independent of $L$. The effect of $h$ is therefore limited by the bound \eqref{eq:Elitzur_bound} due to the presence of local symmetry.  
The crucial point is that
one eventually wants to take
the thermodynamic limit
$\lim_{h \to 0}
   \lim_{L \to \infty}  \langle \hat{\mathcal{L}_0} \rangle$, with
the limit $L \to \infty$  taken before
$h \to 0$.

For a spin system, if $\hat{\mathcal{L}}_0$
is local, then it involves a finite number of spins and its maximum eigenvalue $||\hat{\mathcal{L}}_0||$ is finite. As a consequence, $\langle \hat{\mathcal{L}}_0 \rangle$ goes to zero  for
$h \to 0$,
constituting Elitzur's result. Notice that, if
$\hat{\mathcal{U}}$ were a global symmetry,
then $K$, the number of terms affected by the transformation,  would be infinite in the thermodynamic limit,
invalidating the vanishing of  $\langle \hat{\mathcal{L}}_0 \rangle$.

Moreover, for a bosonic system at finite density and with repulsive interactions, high occupation numbers are exponentially suppressed, and one can artificially set a number cutoff, so that $||\hat{\mathcal{L}}_0||$ is effectively finite.
Instead, in the limit of infinite density, it is not possible to bound
$||\hat{\mathcal{L}}_0||$, and Elitzur's theorem is circumvented.
In other words, one needs to perform an additional limit, where the number of local degrees of freedom is sent to infinity, and this enables a local spontaneous symmetry breaking; in this case,  mean-field theory can be a reasonable approach.

In our case,
the $\hat{\mathcal{L}}_{z,j}$ operator plays the role of  $\hat{\mathcal{L}}_{j}$,
and $\hat{\mathcal{U}}$ can be chosen either as $\hat{\mathcal{U}}_{-1,0}$ or $\hat{\mathcal{U}}_{0,1}$; finally, $K=2$ plaquettes are affected by $\hat{\mathcal{U}}$.
In other words, the angular momentum operator and its projection are not invariant under the local gauge transformation and, more specifically, they anti-commute with the local symmetry (when acting on the plaquette of interest). 
As a consequence, Elitzur's theorem applies, to yield

\begin{equation}
\langle  \hat{\mathcal{L}}_{z,\plaqindex} \rangle   = 0
\ \ \ {\rm and} \ \ \
    \langle  \hat{L}_{z,\plaqindex}  \rangle   = 0.
\end{equation}

\section{Mean-field theory and Bogoliubov modes}
\label{ap:MF}

In this appendix, we discuss the form of the ground states in mean-field (MF) theory, calculate the Bogoliubov modes and derive the sound velocity used in Fig. \ref{fig:MF_evol}. 
For a very large number of bosons per plaquette $\filling \gg 1$ and very weak interaction $g \equiv U \filling \ll J$, a MF approximation is justified, replacing all the operators by complex numbers $\hat{\alpha}_{\sigma, \plaqindex} \rightarrow \alpha_{\sigma, \plaqindex} \equiv \sqrt{\filling \rho_{\sigma, \plaqindex}}\e^{i\theta_{\sigma, \plaqindex}}$, with $\rho_{\sigma \plaqindex}$ and $\theta_{\sigma, \plaqindex}$ being respectively the intensity and the phase of the field at site $\sigma$ of the $\plaqindex$-th plaquette. 
Notice that with this normalization $\frac{1}{L}\sum_{\sigma \plaqindex} \rho_{\sigma, \plaqindex} = 1$.
The Gross-Pitaevskii equation associated with Eq. \eqref{eq:H} is written
\begin{equation}
    \begin{split}
        & i \partial_t a_\plaqindex = -J' d_{\plaqindex-1} - J(-b_\plaqindex + c_\plaqindex) + U|a_\plaqindex|^2 a_\plaqindex, \\
        & i \partial_t b_\plaqindex = - J(-a_\plaqindex + d_\plaqindex) + U|b_\plaqindex|^2 b_\plaqindex, \\
        & i \partial_t c_\plaqindex = - J(a_\plaqindex + d_\plaqindex) + U|c_\plaqindex|^2 c_\plaqindex, \\
        & i \partial_t d_\plaqindex = -J' a_{\plaqindex+1} - J(b_\plaqindex + c_\plaqindex) + U|d_\plaqindex|^2 d_\plaqindex.
    \end{split}
    \label{eq:GP_eq}
\end{equation}
The results of Fig.~\ref{fig:MF_evol}
are obtained by numerically solving these equations with a split-step method.

The symmetries of the Hamiltonian and the form of the single-particle Bloch states entail that one of the many MF ground states has the form
\begin{equation}
    \begin{pmatrix}
        a_\plaqindex  \\
       b_\plaqindex \\
        c_\plaqindex \\
        d_\plaqindex 
    \end{pmatrix}
    = \sqrt{\filling} e^{i \plaqindex\pi/2} \begin{pmatrix}
    \sqrt{\rho_A}e^{-i\pi/2} \\ 
    \sqrt{1/2 - \rho_A} e^{i\pi/4} \\
    \sqrt{1/2 - \rho_A} e^{-i\pi/4} \\
    \sqrt{\rho_A}
    \end{pmatrix}.
    \label{eq:MF_approx}
\end{equation}
As explained in Sec. \ref{subsec:MF}, the other MF ground states can be found by repeated application of the local symmetry operations, which correspond to switching the chirality of a pair of plaquettes. 
Using Eq. \eqref{eq:MF_approx}, the minimization of the MF energy functional yields  the following fourth order equation  for $x \equiv \rho_A/\filling$,
\begin{equation}
    \begin{split}
        &  4g^2x^4 - 4g(J' + g)x^3 + \Big(8 J^2 + J'^2 + 3J'g + \frac{5}{4}g^2 \Big)x^2 \\
        &  - \frac{1}{2}\Big( 8J^2 + J'^2 + J'g + \frac{g^2}{4}\Big)x + \frac{1}{2}J^2 = 0.
    \end{split}
    \label{eq:4th_order_MF}
\end{equation}

\begin{figure}[t]
    \centering
    \includegraphics[scale=0.4]{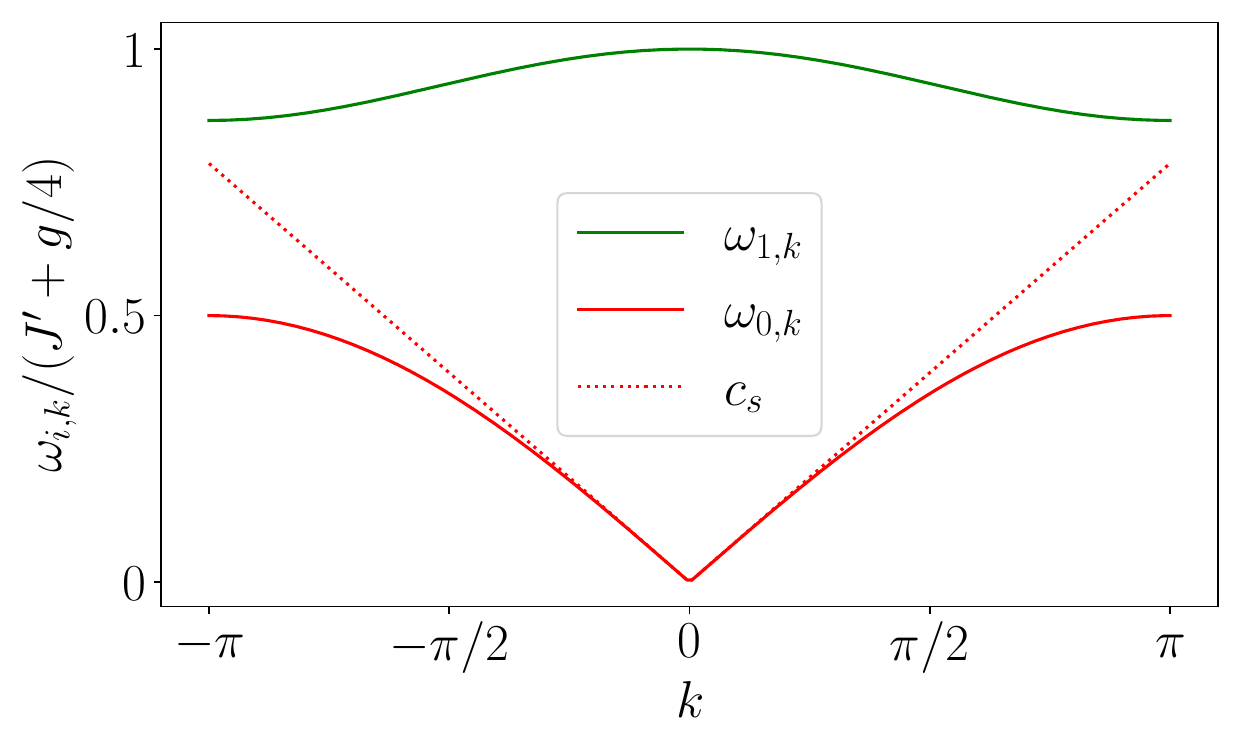}
    \caption{Bogoliubov spectrum, consisting of the acoustic branch $\omega_{0,k}$ (red solid line) of sound velocity $c_s$ (red dots), and of the gapped mode $\omega_{1,k}$ (green line), as   calculated from Eq. \eqref{eq:bog_modes}. Here $g/J' = 4$.
    }
    \label{fig:bog_spectrum}
\end{figure}

We now investigate the low energy collective modes of the system in the Bogoliubov approach.
In order to simplify the analysis, we  consider the regime
$J' \ll J$
and  treat the projected Creutz ladder model of Eq.~(\ref{eq:Hproj_total}) in the MF approximation. 
It is straightforward to verify that one MF ground state reads
$\langle \hat{\beta}_{+,\plaqindex} \rangle = \e^{i\plaqindex\pi/2}\sqrt{\plaqindex}$ and $\langle \hat{\beta}_{-,\plaqindex} \rangle = 0$.
Considering the fluctuations $\delta \hat{\beta}_{\pm,\plaqindex} = \frac{1}{\sqrt{L}} \sum_k \delta \hat{\beta}_{\pm,k} e^{ik\plaqindex}$ and expanding the Hamiltonian to 2nd order yields
\begin{equation}
    \begin{split}
         \hat{H}^{\text{proj}}_{\rm Bogo} = \sum_k & \Bigg[ \xi_1(k) \delta \hat{\beta}^\dagger_{+,k} \delta \hat{\beta}_{+,k} + \xi_2(k) \delta \hat{\beta}^\dagger_{-,k} \delta \hat{\beta}_{-,k} \\
         & + \alpha(k) \big( \delta \hat{\beta}^\dagger_{+,k} \delta \hat{\beta}_{-,k} + \delta \hat{\beta}^\dagger_{-,k} \delta \hat{\beta}_{+,k} \big) \\
         & + \frac{g}{8} \big( \delta \hat{\beta}^\dagger_{+,k} \delta \hat{\beta}^\dagger_{+,-k} +  \delta \hat{\beta}_{+,k} \delta \hat{\beta}_{+,-k} \big) \Bigg],
    \end{split}
\end{equation}
with $\xi_1(k) = \frac{J'}{2}(1 - \cos{k}) + \frac{g}{4} $, $\xi_2(k) = \frac{J'}{2}(1 + \cos{k}) + \frac{g}{4} $ and $\alpha(k) = -\frac{J'}{2} \sin{k}$. Diagonalization of this Hamiltonian leads to the following equation for the Bogoliubov modes
\begin{equation}
\begin{split}
     & \omega_k^4 - \omega_k^2 \Big(2\alpha^2 + \xi_1^2 + \xi_2^2 - \frac{g^2}{16} \Big) \\
     & + \big(\xi_1 \xi_2 - \alpha^2\big)^2 - \frac{\xi_2^2g^2}{16} = 0, 
\end{split}
\label{eq:bog_modes}
\end{equation}
consisting of the two (positive energy)  bands $\omega_{0,k}$ and $\omega_{1,k}$ displayed in Fig.~\ref{fig:bog_spectrum}. At low momentum, $k \rightarrow 0$, the Goldstone branch has dispersion
$\omega_{0,k} \simeq  c_s |k|$, with
\begin{equation}
    c_s = \frac{1}{2\sqrt{2}}  \sqrt{\frac{J'}{g + 4J'}}g 
    \label{speed_of_sound}
\end{equation}
being the speed of sound.

\section{Validity of the projected model}
\label{ap:Projection}

\begin{figure}[h!]
    \centering
    \includegraphics[scale=0.5]{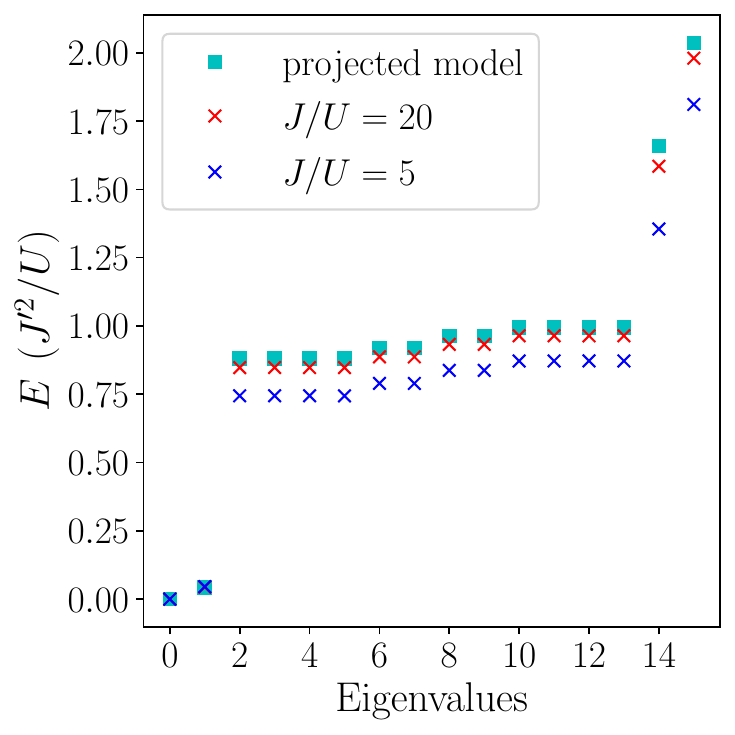}
    \caption{The spectra calculated via exact diagonalization of the full model of Eq.~\eqref{eq:H} for different values of $J/U$ is compared to the spectrum of the projected model of Eq.~\eqref{eq:Hproj_total}, derived assuming $J \gg U$. Here we consider 4 unit cells (corresponding to 16 sites for the full model and 8 for its projection) filled with 4 bosons ($\nu = 1$), for $U/J' = 20$.
    }
    \label{fig:proj_validity}
\end{figure}

In this appendix, we study the full model without any approximation and verified that the results predicted using the projection qualitatively hold for reasonable values of $J > U, J'$. We calculate the spectrum of the full Hamiltonian \eqref{eq:H} and compare it to the spectrum of the projected model \eqref{eq:Hproj_total} in Fig. \ref{fig:proj_validity}. Notice that we choose the same $U/J'$ here as the ones used in Fig. \ref{fig:E_spin_mapping}.a.

\section{Extension to higher dimensions}
\label{ap:2D}

\begin{figure}[h!]
\centering
\includegraphics[scale=0.73]{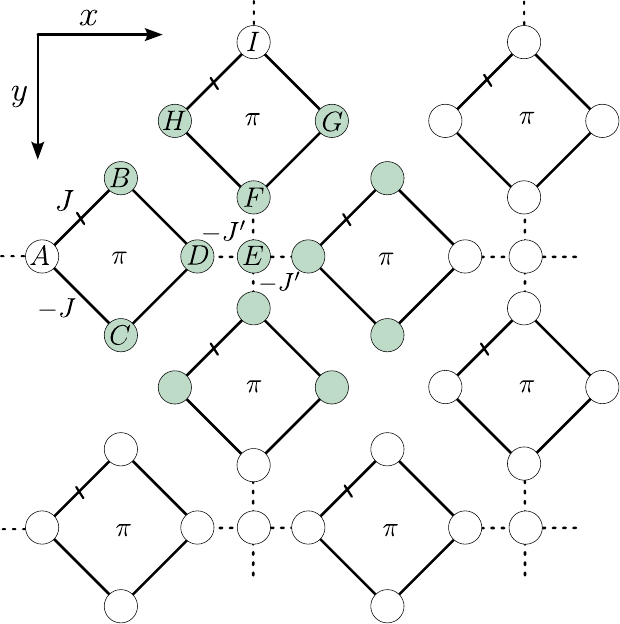}
\caption{Sketch of the two-dimensional version of our model, consisting of $\pi$-flux plaquettes connected with weak links $J'$. The $(x,y)$ unit cell is composed of the nine sites, labeled from $A$ to $I$; three other unit cells, at $(x+1,y)$, $(x,y+1)$ and $(x+1,y+1)$, are plotted here. A set of local symmetries can be defined in this system. Namely, the local symmetry transformation $\hat{\mathcal{U}}_{(x,y)}$, mathematically defined in Eq.~\eqref{eq:loc_symm_2D}, involves the thirteen sites colored in green.
}
\label{fig:2D_model}
\end{figure}

Inspired from our 1D chain connecting plaquettes with weak links $J'$, we provide a natural extension to higher dimensions sketched in Fig. \ref{fig:2D_model}, which is of interest for future works. This 2D flat band model features two $\pi$-flux plaquettes per unit cell, connected by $J'$ links with an additional $E$ site. Notice that a 3D version is easily constructed by connecting additional plaquettes to this specific site along the $z$ direction.

\noindent The associated single-particle 2D Hamiltonian is written as

\begin{equation}
\begin{split}
    \hat{H}_{2D} = \sum_{x,y} \Bigg[ -J \bigg(
     & -\hat{a}_{x,y}^\dagger \hat{b}_{x,y} + \hat{a}_{x,y}^\dagger \hat{c}_{x,y} + \hat{b}_{x,y}^\dagger \hat{d}_{x,y} \\
    & + \hat{c}_{x,y}^\dagger \hat{d}_{x,y} + \hat{f}_{x,y}^\dagger \hat{g}_{x,y} + \hat{f}_{x,y}^\dagger \hat{h}_{x,y} \\
    & + \hat{g}_{x,y}^\dagger \hat{i}_{x,y} - \hat{h}_{x,y}^\dagger \hat{i}_{x,y} \bigg) \\
    & \hspace{-0.9cm} - J' \bigg( \hat{d}_{x,y}^\dagger \hat{e}_{x,y} + \hat{e}_{x,y}^\dagger \hat{f}_{x,y} + 
    \hat{e}_{x,y}^\dagger \hat{a}_{x+1,y} \\ & + \hat{e}_{x,y}^\dagger \hat{i}_{x,y+1} \bigg) + H.C. \Bigg].     
\end{split}
\label{eq:H_2D}
\end{equation}

One can verify that, upon diagonalization of \eqref{eq:H_2D}, the dispersion relation displays nine perfectly flat bands, related to the presence of an extensive set of local symmetries. These can be built from the operators $\hat{\mathcal{U}}_{(x,y)}$, which have a nontrivial action over the four plaquettes connected by the $E$ site of the $(x,y)$ unit cell, defined by
\begin{equation}
    \begin{split}
        & \hat{\mathcal{U}}_{(x,y)}^\dagger \, \hat{b}_{x,y} \, \hat{\mathcal{U}}_{(x,y)} = -\hat{c}_{x,y}, \\
        & \hat{\mathcal{U}}_{(x,y)}^\dagger \, \hat{c}_{x,y} \, \hat{\mathcal{U}}_{(x,y)} = -\hat{b}_{x,y}, \\
        & \hat{\mathcal{U}}_{(x,y)}^\dagger \, \hat{d}_{x,y} \, \hat{\mathcal{U}}_{(x,y)} = -\hat{d}_{x,y}, \\
        & \hat{\mathcal{U}}_{(x,y)}^\dagger \, \hat{e}_{x,y} \, \hat{\mathcal{U}}_{(x,y)} = -\hat{e}_{x,y}, \\
        & \hat{\mathcal{U}}_{(x,y)}^\dagger \, \hat{f}_{x,y} \, \hat{\mathcal{U}}_{(x,y)} = -\hat{f}_{x,y}, \\
        & \hat{\mathcal{U}}_{(x,y)}^\dagger \, \hat{g}_{x,y} \, \hat{\mathcal{U}}_{(x,y)} = -\hat{h}_{x,y}, \\
        & \hat{\mathcal{U}}_{(x,y)}^\dagger \, \hat{h}_{x,y} \, \hat{\mathcal{U}}_{(x,y)} = -\hat{g}_{x,y}, \\
        & \\
        & \hat{\mathcal{U}}_{(x,y)}^\dagger \, \hat{a}_{x+1,y} \, \hat{\mathcal{U}}_{(x,y)} = -\hat{a}_{x+1,y}, \\
        & \hat{\mathcal{U}}_{(x,y)}^\dagger \, \hat{b}_{x+1,y} \, \hat{\mathcal{U}}_{(x,y)} = \hat{c}_{x+1,y}, \\
        & \hat{\mathcal{U}}_{(x,y)}^\dagger \, \hat{c}_{x+1,y} \, \hat{\mathcal{U}}_{(x,y)} = \hat{b}_{x+1,y}, \\
        & \\
        & \hat{\mathcal{U}}_{(x,y)}^\dagger \, \hat{i}_{x,y+1} \, \hat{\mathcal{U}}_{(x,y)} = -\hat{i}_{x,y+1}, \\
        & \hat{\mathcal{U}}_{(x,y)}^\dagger \, \hat{h}_{x,y+1} \, \hat{\mathcal{U}}_{(x,y)} = \hat{g}_{x,y+1}, \\
        & \hat{\mathcal{U}}_{(x,y)}^\dagger \, \hat{g}_{x,y+1} \, \hat{\mathcal{U}}_{(x,y)} = \hat{h}_{x,y+1}, \\
    \end{split}
    \label{eq:loc_symm_2D}
\end{equation}
while the action of $\hat{\mathcal{U}}_{(x,y)}$ is trivial on the other sites. \\

We expect the results presented in the main text to qualitatively translate to higher dimensions. 
In particular, Elitzur's theorem prevents the establishment of any long-range chiral order. 
Technical tools such as band projections and effective spin models can be applied. 
Finally, the self-pinning phenomenon is also in principle possible, but needs to be investigated with quantitative methods.


~

~

\bibliography{bibliography.bib}

\clearpage

\end{document}